\documentclass[aps,prb,twocolumn,groupedaddress,showpacs,superscriptaddress,amssymb,amsmath,noeprint]{revtex4-2}
\usepackage{graphicx}
\usepackage[english]{babel}
\usepackage{dcolumn}
\usepackage{bm}
\usepackage{hyperref}
\usepackage{physics}
\usepackage{float}
\usepackage{cleveref}
\usepackage{multirow}
\usepackage{braket}
\usepackage{mathtools}
\usepackage[normalem]{ulem} 
\hypersetup{
    colorlinks=true,
    linkcolor=blue,
    urlcolor=blue,
    citecolor=blue
}
\usepackage{xcolor}
\usepackage{comment}

\newcommand{\la}{\langle}
\newcommand{\ra}{\rangle}

\usepackage{color}
\usepackage{enumitem}
\usepackage[utf8]{inputenc}
\usepackage{graphicx}
\usepackage{tabularx}
\usepackage{xcolor}
\usepackage{amsmath}
\usepackage{dcolumn}
\usepackage{hyperref}
\usepackage{bm}
\usepackage{epsf}
\usepackage{braket}
\usepackage{tensor}
\usepackage{soul}

\begin{document}
\newcolumntype{M}[1]{>{\centering\arraybackslash}m{#1}}

\title{Full counting statistics for boundary driven transport in presence of correlated gain and loss channels}

\author{Katha Ganguly}
\email{katha.ganguly@students.iiserpune.ac.in}
\affiliation{Department of Physics, Indian Institute of Science Education and Research Pune, Dr. Homi Bhabha Road, Ward No. 8, NCL Colony, Pashan, Pune, Maharashtra 411008, India}

\author{Bijay Kumar Agarwalla}
\email{bijay@iiserpune.ac.in}
\affiliation{Department of Physics, Indian Institute of Science Education and Research Pune, Dr. Homi Bhabha Road, Ward No. 8, NCL Colony, Pashan, Pune, Maharashtra 411008, India}

\date{\today}

\begin{abstract}
One of the major advances of quantum technology is the engineering of complex quantum channels in lattice systems that paves the way for a variety of novel non-equilibrium phenomena. For a boundary driven lattice with such engineered quantum channels, the analysis of the full counting statistics of current across boundaries has received limited attention. In this work, we consider a boundary driven free fermionic lattice with carefully engineered correlated gain and loss channels and
obtain the cumulant generating function of the steady-state particle current. We also discuss the limit for simplifying the correlated gain-loss channel to a local gain-loss channel and
obtain the average current and its fluctuation in such cases. Generally, in the presence of gain-loss, the current statistics are different at the two ends of the lattice. Hence, for both local and correlated gain-loss, we devise the conditions for which the statistics can coincide, giving rise to a $\mathcal{PT}$ symmetric balanced gain-loss scenario. A striking difference between the correlated gain-loss and their local counterpart is the emergence of nonreciprocity in the system and we observe that it has a dramatic impact in the current as well as fluctuations. Our work therefore provides interesting insights about the importance of engineered dissipators in boundary driven systems.
\end{abstract}

\maketitle

\section{Introduction}
Boundary driven quantum transport in small-scale, low-dimensional systems has always been a thriving field of research~\cite{Dario_Landi2022,Dhar2006,PhysRevB.75.195110,abhishek2008,subdiffusive_long,PhysRevB.102.224512,PEREIRA2020126864,PhysRevB.98.075421,Agarwalla_2017,PhysRevA.105.013307,PhysRevB.81.085330,roy2023nonequilibrium,PhysRevLett.106.220601,Ljubotina2017,PhysRevB.88.205135,Bulchandani_2021}. Over the past decade, there has been immense progress in its theoretical understanding which ranges from classifying transport regimes~\cite{Purkayastha_2019,PhysRevB.97.174206,Dario_Landi2022} to observation of diverse nonequilibrium phenomena~\cite{PhysRevLett.127.240601,PhysRevB.102.245433,PhysRevB.102.195142,PhysRevB.96.180204} in non-interacting as well as interacting quantum many-body systems~\cite{PhysRevX.13.011033,Chien2015QuantumTI,PhysRevB.86.125118,RevModPhys.93.025003,PhysRevB.99.241113,Sieberer_2016,PhysRevB.105.134203}. Such a surge in the study of quantum transport is also escalated by the advancement of various quantum technological platforms such as superconducting qubits, trapped ions, optical lattices, neutral atoms, etc.~\cite{Schäfer2020,Bloch2012,Rudolf2013,Blatt2012,Barredo2016,Jepsen2020,Greiner2002,Jördens2008,Atala2014,PhysRevA.85.041601,Rajibul2010}. These platforms not only realize the conventional quantum transport setup with system and reservoirs, rather enable the engineering of complex quantum channels with high controllability and scalability~\cite{sebastian2008,nature2009,Kater2022,Korenblit_2012}. 

Theoretically, many such engineered quantum channels are often modeled by the Gorini - Kossakowski - Sudarshan - Lindblad (GKSL) Quantum Master Equation ~\cite{L1976,VG1976,BPOQS}. A few examples of such dissipators/channels in the lattice setup include local and correlated gain and loss~\cite{Giamarchi2022,Giamarchi2023,loss2024,schiro2025}, quasiparticle dephasing~\cite{wang2023,yusipov2017}, symmetric exclusion process~\cite{fujimoto2022,Denis2023}, etc. Recent theoretical studies have shown that these dissipators give rise to many intriguing phenomena such as emergence of steady-state localization through engineered dissipation~\cite{Jiangbin2025,wang2024,Vershinina_2017,Denisov2018}, anomalous transport driven by quasiparticle dephasing~\cite{wang2023} the absence of measurement induced phase transition and non-hermitian skin effect~\cite{JieRen2024}. Further, dissipation induced
edge charge accumulation, emergence of chirality, directionality of current etc. have also catalyzed the growing interest of engineered dissipation~\cite{schiro2025}. These dissipators which are modeled phenomenologically via GKSL equation, may not always have direct correspondence to conventional particle reservoirs. Instead their implementation and design require careful bath engineering using laser fields, photonic cavity, optical pumping, coherent driving and feedback control etc.~\cite{Hauke2018,Blatt2019,Cai_2022,Damanet_2019}. Hence it is not always feasible to draw a connection between such engineered  dissipators and microscopic open quantum system models defined by a system–reservoir Hamiltonian~\cite{BPOQS,carmichael2009}. Furthermore, the engineered dissipators are often implemented with high tunability and they are more controllable than natural reservoirs. 
Hence dissipation engineering can serve as a fertile ground for efficiently manipulating quantum systems to obtain desired functionalities~\cite{Mebrahtu2013,Andrew2019,Horodecki2024,ClerkAashish2013,Clerk2015,Clerk2022,Andreas2018,Andreas2023,PhysRevLett.88.094302,PhysRevLett.93.184301,dutta2025}. 

Despite the profound impact of engineered dissipation on quantum transport, the understanding of the full counting statistics of currents in such systems has not yet reached the same level of maturity. A general analytical framework for computing the cumulant generating function of particle current in the presence of engineered dissipation is still lacking. Even for a simple boundary-driven free-fermionic lattice setup which evolves under a generic quadratic Lindblad equation, computation of the full current statistics is nontrivial unless a well established connection to microscopic Hamiltonian models exists. Although such simple noninteracting systems offer exact solution of current by using correlation matrix, computing fluctuations and higher cumulants requires tedious matrix manipulation~\cite{Massi_RevModPhys.81.1665,Landi_FCS2024,Saito2011}.

In this direction, in Ref.~\cite{uchino2023particle},  the cumulant generating function for boundary current is obtained for a quantum point contact and a single quantum dot setup connected to a lossy channel. Furthermore, recently, in Ref.~\cite{Landi_FCS2024}, a numerical scheme is proposed to calculate the full distribution of boundary currents in systems governed by GKSL equation. However, for larger system sizes, such numerics can become extremely challenging. Hence, a general analytical expression of cumulant generating function is highly advantageous for systems governed by GKSL type engineered dissipators.

The agenda of our work is to obtain the full counting statistics (FCS) of steady-state current in boundary driven free fermionic lattice which is further subjected correlated particle gain and loss channels in the bulk. For simplicity, we have modeled the boundary drive by local dissipator although it can be related to ideal fermionic reservoirs in high temperature and high chemical potential limit~\cite{Giamarchi_Lindblad}. Note that, the analysis presented here can be extended to higher dimensional
noninteracting lattices, other kinds of complex quadratic
dissipators, as well as for quadratic bosonic setups. Our approach can also be adapted in a hybrid setup that is modeled as a GKSL equation based engineered dissipations and Hamiltonian based boundary drives~\cite{Giamarchi_Lindblad,loss2024,Giamarchi2022,Giamarchi2023} and thus, can capture non-markovian as well as strong coupling effects on current statistics.

The paper is organized as follows. In Sec.~\ref{sec:setup}, we introduce our setup of the fermionic lattice with boundary drive and correlated particle loss and gain in the bulk. In Sec.~\ref{sec:boundary}, we first derive the cumulant generating function for the particle current by following Keldysh path integral formalism without gain-loss channels. This section sets the stage for developing the path integral formalism for complex dissipators, which is also a reason behind our choice of boundary drives modeled by GKSL equation. In Sec.~\ref{sec:correlated}, we discuss the counting statistics in two cases - local gain-loss channels in Sec.~\ref{subsec:localloss}, and correlated gain-loss channels in Sec.~\ref{subsec:correlated_gainloss}.  We  then analyze the current and noise from the CGF for the setup. In Sec.~\ref{summary}, we summarize our main findings with a future outlook. We provide certain technical details in the appendices.

\section{Fermionic transport setup and counting statistics} 
\label{sec:setup}

\begin{figure}
    \centering
    \includegraphics[width=8.6cm,height=3.3cm]{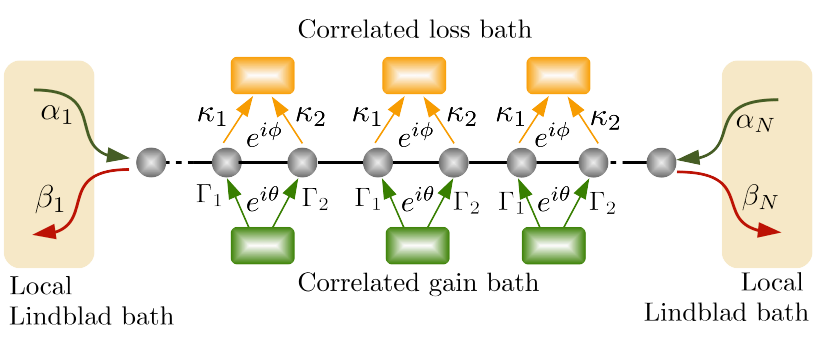}
    \caption{Schematic of the setup considered in this work: The system of interest is a one-dimensional fermionic chain with boundary injection and extraction of fermions from $1$st and $N$-th site with rates $\alpha_1$, $\beta_1$ and $\alpha_N$, $\beta_N$, respectively. Different values of $\alpha_1$, $\beta_1$, and $\alpha_N$, $\beta_N$ drive the system out of equilibrium and produce a steady current in the chain. The lattice can be further subjected to correlated particle loss or gain with gain strength $\Gamma_1$, $\Gamma_2$ and loss strength $\kappa_1$, $\kappa_2$, respectively (see Eq.~\eqref{bulk_dissipator} and discussions below that). The correlated loss (gain) can be reduced to local loss (gain) by putting $\Gamma_2=0$ ($\kappa_2=0$). We first discuss the case with $\Gamma_1=\Gamma_2=\kappa_1=\kappa_2=0$ in Sec.~\ref{sec:boundary} and later discuss the results for correlated gain and loss in Sec.~\ref{sec:correlated}.}
    \label{setup}
\end{figure}
We consider a generic one-dimensional non-interacting fermionic lattice that is driven out of equilibrium by local injection and extraction of particles at its boundaries. In addition to these boundary drives, the setup is subjected to correlated loss and gain channels at each site of the lattice system. The schematic of our setup is shown in Fig.~\ref{setup}.  The time-evolution of the system is described by the Markovian GKSL Quantum Master Equation and is given as,
\begin{align}
    \frac{\partial \rho}{\partial t}=\mathcal{L}\rho \equiv -i\big[H,\rho\big]+\mathcal{D}_{1}[\rho]+\mathcal{D}_N[\rho]+\mathcal{D}_{\rm bulk}[\rho],
    \label{Lindblad_eq}
\end{align}
where, the Liouville superoperator $\mathcal{L}$ has two contributions: the first term represents the unitary evolution which is governed by the Hamiltonian $H$ of the lattice and is considered here as $H=\sum_{i,j=1}^{N} h_{ij}c^{\dagger}_i c_j$ with $h_{ij}$ being the single particle Hamiltonian. Here $c^{\dagger}_{i}$ ($c_{i}$) are the creation (annihilation) operator of fermions at the site $i$. The injection and extraction of fermions at the boundaries (i.e., from the $1$st and $N$-th site of the chain of size $N$) serve as the dissipative contributions to the evolution and is denoted by $\mathcal{D}_1[\rho]$ and $\mathcal{D}_N[\rho]$, respectively. These local injection and extraction (see Fig.~\ref{setup}) are modeled using the local jump operators $c_i$ and $c_i^{\dagger}$, with $i=1$ and $N$, and it is given by, 
\begin{align}
\mathcal{D}_{i}[\rho]=\alpha_{i}\big[2 c^{\dagger}_{i}\rho c_{i}-\{c_{i}c^{\dagger}_{i},\rho\} \big]+\beta_{i}\big[2 c_{i}\rho  &c^{\dagger}_{i}-\{c^{\dagger}_{i}c_{i},\rho\} \big], \nonumber\\& \quad\,\, i\in 1, N
\end{align}
Here, $\alpha_i$ and $\beta_i$ $(i=1, N)$ are the rate of injection and extraction, respectively. A mismatch between the value of $\alpha_i$ and $\beta_i$ results in a non-equilibrium scenario in which a particle current gets generated between the two ends.
Hence these rates $(\alpha_i,\beta_i)$ at any boundary site suffice as a reservoir which we name as the Lindblad reservoir in our analysis. Such Lindblad reservoirs are often identified with ideal fermionic reservoirs in the high temperature and high chemical potential limit keeping their ratio fixed~\cite{Giamarchi_Lindblad}. The dissipator $\mathcal{D}_{\rm bulk}[\rho]$ in Eq.~\eqref{Lindblad_eq} represents the correlated loss and gain channels throughout the bulk of the lattice, as shown schematically in Fig.~\ref{setup}. This dissipator takes the form,
\begin{align}
    \mathcal{D}_{\rm bulk}[\rho]=\sum_{i=1}^{N-1} \big[2 A_{i}\rho A^{\dagger}_{i}-&\{A^{\dagger}_{i}A_{i},\rho\} \big]\nonumber\\&+\big[2 B_{i}\rho B^{\dagger}_{i}-\{B^{\dagger}_{i}B_{i},\rho\} \big], \label{bulk_dissipator}
\end{align}
with $A_i$ and $B_i$ being 
\begin{eqnarray}
A_i &=&\sqrt{\Gamma_1}c^{\dagger}_i+\sqrt{\Gamma_2}\,e^{i\theta}c^{\dagger}_{i+1}, \nonumber \\
B_i &=& \sqrt{\kappa_1}c_i+\sqrt{\kappa_2}\,e^{i\phi}c_{i+1}. \label{eq:AB_def}
\end{eqnarray} 
Unlike the boundary drive, the bulk dissipator is modeled using the superposition of adjacent local operators $c_i$ and $c_{i+1}$, thereby developing additional correlation between neighboring sites through the action of the dissipator. Such correlated dissipator can be designed in quantum technological platforms by careful reservoir engineering~\cite{Clerk2015,Clerk2022,Andreas2018,Andreas2023}.
Note that, in the special limit, it can represent local gain-loss at each lattice site $i$. This corresponds to choosing $\Gamma_2= \kappa_2=0$. Impact of local loss has been extensively explored in recent times in various contexts such as particle transport~\cite{Giamarchi2022,Giamarchi2023,uchino2023particle,uchino2022loss,Huang2023,loss2024}, entanglement entropy dynamics~\cite{SchiroLoss2024}, engineering novel out-of-equilibrium phenomena~\cite{wolff2020,sebastian2020,Shovan2021}, etc.

We are interested here in computing the full statistics of the integrated particle current that is flowing across the system from the left to the right boundary. The integrated current is defined as
$\mathcal{N}_{L}=\int_{t_0}^{t_f} dt' \, \mathcal{I}_{L}(t')$,
where $t_0$ and $t_f$ refers to the initial and the final time of the counting process, respectively. $\mathcal{N}_L$ refers to the total number of particles received by the $1$st site in the time interval $[t_0,t_f]$ and this quantity, in general, is stochastic in nature. 
To keep track of the statistics of the current from the left reservoir, we
introduce a counting field $\lambda$ corresponding to the injection and extraction of particles from the left reservoir, i.e., in the jump terms of the dissipator $D_1[\rho]$. The counting field dressed GKSL quantum master equation is given as,
\begin{align}
    \frac{\partial \rho}{\partial t}=\mathcal{L}_{\lambda}\rho=-i\big[H,\rho\big]&+\mathcal{D}^{\lambda}_{1}[\rho]+\mathcal{D}_N[\rho]+\mathcal{D}_{\rm bulk}[\rho],\label{Lindblad_eq_counting_field}
\end{align}
where $\mathcal{L}_{\lambda}$ is the dressed Liouvillian and the dressed dissipator $\mathcal{D}^{\lambda}_{1}[\rho]$ is given as,
 \begin{align}
 \mathcal{D}^{\lambda}_{1}[\rho]=\alpha_{1}\big[2e^{i\lambda} c^{\dagger}_{1}\rho c_{1} & -\{c_{1}c^{\dagger}_{1},\rho\} \big]\nonumber\\&+\beta_{1}\big[2e^{-i\lambda}c_{1}\rho c^{\dagger}_{1}-\{c^{\dagger}_{1}c_{1},\rho\} \big]. \label{eq:dressed_D}
 \end{align}
Note that the $\lambda$ dependence appears only in the jump terms. Starting with Eq.~\eqref{Lindblad_eq_counting_field}, we derive the moment generating function (MGF) 
for the full statistics of the 
integrated current $\mathcal{N}_L$. The MGF $\mathcal{Z}(\lambda)$ is given as~\cite{Landi_FCS2024,Hangi2012},
\begin{align}
\label{eq:gf_def}
    \mathcal{Z}(\lambda)=\mathrm{Tr}\Big[e^{(t_f-t_0)\mathcal{L}_{\lambda}}\,\rho(t_0)\Big],
\end{align}
where $\rho(t_0)$ is the initial density matrix of the lattice system. We derive here compact analytical expressions for the MGF following the path integral formalism~\cite{Sieberer_2016,kamenev2011field}. In what follows, we first derive the particle current statistics without the correlated dissipator (Case 1) in Sec.~\ref{sec:boundary}, and then generalize our study for the correlated scenario (Case 2) in Sec.~\ref{sec:correlated}.

\section{Case 1: Only boundary drives}
\label{sec:boundary}
In this section, we first derive the generating function defined in Eq.~\eqref{eq:gf_def} for the given setup subjected to only boundary drives i.e., we set $\mathcal{D}_{\rm bulk}[\rho]=0$ in Eq.~\eqref{Lindblad_eq_counting_field}. 
We can express $\mathcal{Z}(\lambda)$ following the path integral formalism as~\cite{kamenev2011field,rammer2007quantum,daniel2025,Aashish2023_KFT}
\begin{equation}
    \mathcal{Z}(\lambda)\!=\!\int \! \! \! D\big[\Psi^{*}_{+},\Psi_{+},\Psi^{*}_{-},\Psi_{-}\big]\,e^{i S(\lambda)}\, \big \langle \Psi_{+}(t_0)|\rho(t_0)|\Psi_{-}(t_0)\big \rangle.\label{partition_function}
\end{equation}
Here $\Psi=\big(\psi_1\,\, \psi_2\,\, \psi_3\,\, \cdots, \psi_{N}\big)^{T}$ is a $N$ dimensional vector with each $\psi_i$ being the Grasmann variables corresponding to site $i$ of the lattice. This is introduced using the eigenvalue equation $c_i |\Psi\ra=\psi_i|\Psi\ra$ where $\psi_i$ is the eigenvalue of the local annihilation operator $c_i$ and as the fermionic statistics rely on anticommutation algebra, $\psi_i$ is represented by Grassmann variable. $\Psi^{*}$ represents complex conjugate, the subscript $+$ and $-$ correspond to the forward and backward evolution of the density matrix $\rho$ described by the GKSL equation in Eq.~\eqref{Lindblad_eq_counting_field}. $D[\Psi^{*}_{+},\Psi_{+},\Psi^{*}_{-},\Psi_{-}]$ in Eq.~\eqref{partition_function} represents, 
\begin{align}
    D[\Psi^{*}_{+},\Psi_{+},\Psi^{*}_{-},\Psi_{-}]=\lim_{\delta_t \rightarrow 0}\prod_{n=1}^{\frac{(t_f-t_0)}{\delta_t}}\frac{d\Psi^{*}_{n+}d\Psi_{n+}}{\pi}\frac{d\Psi^{*}_{n-}d\Psi_{n-}}{\pi}.
\end{align}
The counting field dressed Keldysh action $S(\lambda)$ in Eq.~\eqref{partition_function} is given as  [see Appendix.~\ref{app:action} for the details of the derivation],
\begin{align}
S(\lambda)= \int_{t_0}^{t_f} dt \, \Big[\Psi^{*}_{+}& \,\big(i \, \partial_t \big) \, \Psi_{+} - \Psi^{*}_{-}\,  \big(i \, \partial_t \big)\Psi_{-}\, \nonumber\\& - i\,f_{\lambda}(\Psi^{*}_{+},\Psi_{+},\Psi^{*}_{-},\Psi_{-})\Big], \label{action1}
\end{align}
where the term $f_{\lambda}(\Psi^{*}_{+},\Psi_{+},\Psi^{*}_{-},\Psi_{-})$ contains three different contributions,
\begin{widetext}
\begin{align}
\label{action}
f_{\lambda}(\Psi^{*}_{+},\Psi_{+},\Psi^{*}_{-},\Psi_{-})=f_H(\Psi^{*}_{+},\Psi_{+},\Psi^{*}_{-},\Psi_{-})+f^{\lambda}_1(\Psi^{*}_{+},\Psi_{+},\Psi^{*}_{-},\Psi_{-})+f_N(\Psi^{*}_{+},\Psi_{+},\Psi^{*}_{-},\Psi_{-}).
\end{align}
The contribution due to the unitary part  i.e., $f_H(\Psi^{*}_{+},\Psi_{+},\Psi^{*}_{-},\Psi_{-})$ is given by,
\begin{align}
f_H(\Psi^{*}_{+},\Psi_{+},\Psi^{*}_{-},\Psi_{-})\!=\!-i\!\!\sum_{i,j=1}^{N}h_{ij}\big[\psi^{i*}_{+}\psi_{+}^{j}\!-\!\psi_{-}^{i*}\psi_{-}^{j}\big]. \label{fH}
\end{align}
The other two contributions $f^{\lambda}_1(\Psi^{*}_{+},\Psi_{+},\Psi^{*}_{-},\Psi_{-})$ and $f_N(\Psi^{*}_{+},\Psi_{+},\Psi_{-}^{*},\Psi_{-})$ come from the left and right boundary drives and are given by,
    \begin{align}
&f^{\lambda}_1(\Psi^{*}_{+},\Psi_{+},\Psi^{*}_{-},\Psi_{-})=\alpha_1[2e^{i\lambda}\psi^{1*}_{+}\psi_{-}^{1}+\psi^{1*}_{+}\psi_{+}^{1}+\psi^{1*}_{-}\psi_{-}^{1}]+\beta_1[2e^{-i\lambda}\psi^{1}_{+}\psi_{-}^{1*}-\psi^{1*}_{+}\psi_{+}^{1}-\psi^{1*}_{-}\psi_{-}^{1}],\label{f1}\\
    &f_N(\Psi_{+}^{*},\Psi_{+},\Psi_{-}^{*},\Psi_{-})=\alpha_N[2\psi^{N*}_{+}\psi_{-}^{N}+\psi^{N*}_{+}\psi_{+}^{N}+\psi^{N*}_{-}\psi_{-}^{N}]+\beta_N[2\psi^{N}_{+}\psi_{-}^{N*}-\psi^{N*}_{+}\psi_{+}^{N}-\psi^{N*}_{-}\psi_{-}^{N}]\label{fN}.
\end{align}
These are obtained by expressing the Liouvillian superoperator defined in Eq.~\eqref{Lindblad_eq} in  the coherent basis $\{|\Psi_{\pm}\ra\}$.
We are interested in the statistics of the steady-state integrated current and hence we consider the long-time limit i.e., $t_0 \to -\infty$ (starting from remote past) and $t_f\to \infty$ (ending at remote future) of Eq.~\eqref{partition_function}. We then write down the action $S(\lambda)$ in a compact form in the frequency domain by performing Fourier transformation. We obtain 
\begin{align}
    S(\lambda)=\int_{-\infty}^{\infty} \frac{d\omega}{2\pi}\, \begin{pmatrix}
        \Psi^{*}_{+}(\omega) & \Psi^{*}_{-}(\omega) \\
    \end{pmatrix} \Big[\mathcal{M}_0(\omega)+\mathcal{M}_{\lambda}\Big]_{2N\times 2N} \begin{pmatrix}
        \Psi_{+}(\omega) \\
        \Psi_{-}(\omega) \\
    \end{pmatrix}, \label{action_matrix_form}
\end{align}
where we organize the elements of the matrix $\mathcal{M}_0(\omega) =\begin{pmatrix}
    A(\omega) & B(\omega)\\
    C(\omega) & D(\omega)
\end{pmatrix} $.
Here $A(\omega)$, $B(\omega)$, $C(\omega)$, and $D(\omega)$ are $N\times N$ block matrices with the components given as,
\begin{align}
&[A(\omega)]_{lm}=\omega\delta_{lm}-h_{lm}+i[\gamma^{t}_{1}+\gamma^{t}_{N}]_{lm},\nonumber\\ 
    &[B(\omega)]_{lm}=-i[\gamma^{<}_{1}+\gamma^{<}_{N}]_{lm},\nonumber\\
   &[C(\omega)]_{lm}=- i[\gamma^{>}_{1}+\gamma^{>}_{N}]_{lm},\nonumber\\
        &[D(\omega)]_{lm}=-\omega\delta_{lm}+h_{lm}+i[\gamma^{t}_{1}+\gamma^{t}_{N}]_{lm}.
\end{align}
All the $\gamma$ matrices represent different self energy components and their forms are given by,
\begin{align}
    \gamma_{j,lm}^{t}=-(\alpha_j-\beta_j)\gamma_{j,lm},\quad\gamma_{j,lm}^{<}=2\alpha_{j}\gamma_{j,lm},\,\,\gamma^{>}_{j,lm}=-2\beta_j\gamma_{j,lm},\,\,\,\,{\rm where,}\quad \gamma_{j,lm}=\delta_{jl}\delta_{jm},\,\,{\rm and}\,\, j\in \{1,N\}\label{eq:self_energy_boundary}
\end{align}
Note that, as the boundary drive is modeled by Markovian Lindblad type reservoirs, all the self energies are independent of the frequency variable $\omega$. The counting field dependent matrix $\mathcal{M}_{\lambda}$ in Eq.~\eqref{action_matrix_form} takes the form $\mathcal{M}_{\lambda}=\begin{pmatrix}
    0 & \gamma_1^{<}(\lambda)\\
    \gamma_1^{>}(\lambda) & 0
\end{pmatrix}$ where the components are given by,
\begin{align}
    \gamma_1^{<}(\lambda)=-i(e^{i\lambda}-1)\,\gamma_1^{<}\,\,\,\,{\rm and}\,\,\,\,\gamma_1^{>}(\lambda)=-i(e^{-i\lambda}-1)\,\gamma_1^{>}.
\end{align}
Note that, as the counting process is performed only for the left reservoir, the matrix $\mathcal{M}_{\lambda}$ involves term that depends only on the left Lindblad reservoir with counting parameter $\lambda$.
\end{widetext}
We next perform the Keldysh rotation~\cite{rammer2007quantum} to the matrices  $\mathcal{M}_0$ and $\mathcal{M}_{\lambda}$ in Eq.~\eqref{action_matrix_form} by the orthogonal matrix $\mathcal{O}=\frac{1}{\sqrt{2}}\begin{pmatrix}
    \mathcal{I} & \mathcal{I} \\
    \mathcal{I} & -\mathcal{I} \\ 
\end{pmatrix}$ where $\mathcal{I}$ being the $N\times N$ identity matrix.  We then obtain the action as
\begin{align}
    S(\lambda)=\int_{-\infty}^{\infty} \frac{d\omega}{2\pi}\, \begin{pmatrix}
        \Psi^{*}_{1} & \Psi^{*}_{2} \\
\end{pmatrix}
\big(\Breve{\mathcal{M}}_{0}(\omega)+\Breve{{\mathcal{M}}}_\lambda\big) \begin{pmatrix}
        \Psi_{1} \\
        \Psi_{2} \\
    \end{pmatrix}, \label{action2}
\end{align}
where $\Psi_1=(\Psi_{+}+\Psi_{-})/\sqrt{2}$ and $\Psi_2=(\Psi_{+}-\Psi_{-})/\sqrt{2}$ are the rotated Grassmann field variables. The rotated matrices are denoted by the breve symbol. By a careful observation of the elements of $\Breve{\mathcal{M}}_{0}(\omega)$, we identify the matrix in terms of the various components of non-equilibrium Green's functions (NEGF) as \cite{Giamarchi_Lindblad}, 
\begin{align}
    &\Breve{\mathcal{M}}_{0}(\omega)=\begin{bmatrix}
        -i\big(\gamma_1^{K}+\gamma_N^{K}\big) & \big[G^{R}(\omega)\big]^{-1} \\
        \big[G^{A}(\omega)\big]^{-1} & 0 \\
    \end{bmatrix},
\end{align}
where the retarded component of the Green's function for the lattice is given as 
\begin{equation}
G^{R}(\omega)=\Big[\omega \mathcal{I}-h+i(\gamma^{R}_{1}+\gamma^{R}_{N})\Big]^{-1} \label{eq:retarded}
\end{equation}
which is related to the advanced component by hermitian conjugate i.e., $G^{A}(\omega)= \Big[G^{R}(\omega)\Big]^{\dagger}$. $\gamma_{i}^{R}$ is the retarded self energy component due to the reservoirs and has the form 
\begin{align}
\gamma_{j,lm}^{R}=\,(\alpha_j+\beta_j)\,\delta_{jl}\,\delta_{jm}, \,\, \quad \, j\in\{1,N\}.\label{eq:ret_selfenergy}
\end{align}
The retarded Green's function is also known as the response function due to its casual structure~\cite{kamenev2011field,rammer2007quantum,bruus2004many}. 
Finally, the Keldysh component of the lattice Green's function is given by
\begin{equation}
G^{K}(\omega)= i\, G^{R}(\omega) \Big[\gamma^{K}_{1}+\gamma^{K}_{N} \Big]  G^{A}(\omega), \label{eq:Keldysh}
\end{equation} 
which is useful to obtain local density profile of fermions within the lattice. $\gamma^{K}_{j}$ is the Keldysh self-energy component and has the form
\begin{align}
\gamma^{K}_{j,lm}=2\,(\alpha_j-\beta_j)\, \delta_{jl} \, \delta_{jm},\,\,{\rm where,}\,\, j\in\{1,N\}.\label{eq:Keldysh_selfenergy}
\end{align}
Next, the counting field dependent matrix $\Breve{{\mathcal{M}}}_{\lambda}$, after the Keldysh rotation, takes the form,
\begin{align}
\Breve{{\mathcal{M}}}_\lambda=\mathcal{O} \, D_{\lambda} \mathcal{O}^{T}=\gamma_1 \begin{bmatrix}
        b-a &\,  a+b \\
        - a- b &\, a-b \\
    \end{bmatrix},\label{eq:mlambda}
\end{align}
where $a=i\,\alpha_{1}\,(e^{i\lambda}-1)$ and $b=i\,\beta_{1}\,(e^{-i\lambda}-1)$. It is worth observing that the Keldysh action defined in Eq.~\eqref{action2} is quadratic in Grassmann variables $\Psi_1$ and $\Psi_2$. Such a quadratic structure is specific to the non-interacting Hamiltonians and linear dissipator in the Lindblad equation, given in Eq.~\eqref{Lindblad_eq_counting_field}. As a result, the integral in Eq.~\eqref{partition_function} representing the MGF $\mathcal{Z}(\lambda)$, contains the exponential of the quadratic action and thereby making it a multivariate Gaussian integral in Grassmann variables.  The solution of such a   Grassmann integral is given as $\int_{-\infty}^{+\infty} \prod_{i=1}^{d} dx_i \exp(-\sum_{m,n}x^*_m  B_{mn}x_n) =\det(B)$ where $B$ is a $d\times d$ matrix. Note that the action given in Eq.~\eqref{action2} is in terms of an integral over continuous frequency variable $\omega$ instead of a form like $\sum_{m,n}x^*_m  B_{mn}x_n$. Hence we discretize the frequency space $\omega$ in units of $2\pi/t_M$, where $t_M=t_f-t_0$ is the total time duration of the counting process. We can therefore write the action as,
\begin{align}
    S(\lambda)\approx \sum_{\omega,\omega'} \overline{\Psi}^*_\omega (\Breve{\mathbb{M}}_{0}+\Breve{\mathbb{M}}_{\lambda})_{\omega,\omega'}\overline{\Psi}_{\omega'} \label{eq:discrete_S}
\end{align}
Note that $\Breve{\mathbb{M}}_{0,\lambda}$ is different from $\Breve{\mathcal{M}}_{0,\lambda}$; the former is an infinite dimensional matrix obtained after discretizing the frequency space from the latter which is just a function of $\omega$. Similarly, $\overline\Psi=(\overline{\Psi}_{c}\quad \overline{\Psi}_q)$ where $\overline{\Psi}_{c,q}$ is different than $\Psi_{c,q}(\omega)$; $\overline{\Psi}_{c,q}$ is an infinite dimensional vector obtained from $\Psi_{c,q}(\omega)$ after discretizing $\omega$. Next, we evaluate the Gaussian integral in Eq.~\eqref{partition_function} and obtain the MGF $\mathcal{Z}(\lambda)$ as
\begin{align}
    \mathcal{Z}(\lambda)=i\mathcal{N}_c\Big({\rm det}_{\omega,j}[\Breve{\mathbb{M}}_0+\Breve{\mathbb{M}}_\lambda]\Big). \label{eq:z_lambda}
\end{align}
Here, the subscripts $\omega,j$ correspond to the discretized frequency index and the system's degree of freedom, respectively. Note that we are interested in the current statistics in the steady-state where the initial state is irrelevant and hence we ignore the dependence on the initial state. $\mathcal{N}_c$ in Eq.~\eqref{eq:z_lambda} is the normalization constant of the MGF. To evaluate $\mathcal{N}_c$ we now use the normalization condition, i.e., $\mathcal{Z}(\lambda=0)=1$ and obtain $\mathcal{N}_c=-i\,{\rm det}_{\omega,j}[\Breve{\mathbb{M}}_0^{-1}]$. Thus, the MGF is obtained as
\begin{align}
    \mathcal{Z}(\lambda)&=\mathrm{det}_{\omega,j}\big[\mathbb{I}+\Breve{\mathbb{M}}_{0}^{-1}\Breve{\mathbb{M}}_{\lambda}\big]. \label{partition_function2}
\end{align}
\begin{widetext}
\noindent Next we obtain the expression of the MGF in terms of NEGF, defined in Eq.~\eqref{eq:retarded} and \eqref{eq:Keldysh}. By further simplification of Eq.~\eqref{partition_function2}, we receive the following expression for the MGF,
    \begin{align}
        \mathcal{Z}(\lambda)=\mathrm{det}_{\omega,j}&\Big[\mathbb{I}+2\,\mathbb{G}^{R}\gamma_N\mathbb{G}^{A}\gamma_1(\alpha_N+\beta_N)\big(\alpha_1(e^{i\lambda}-1)+\beta_1(e^{-i\lambda}-1)\big)\nonumber\\&-2\mathbb{G}^{R}\gamma_N\mathbb{G}^{A}\gamma_1
       (\alpha_N-\beta_N) \big(\alpha_1(e^{i\lambda}-1)-\beta_1(e^{-i\lambda}-1)\big)\Big],\label{partition_function3}
    \end{align}
    where recall that $\gamma_{1,lm}=\delta_{1l}\,\delta_{1m}$ and $\gamma_{N,lm}=\delta_{Nl}\,\delta_{Nm}$. The corresponding cumulant generating function (CGF) is given as,
    \begin{align}
        \Tilde{\mathcal{G}}_1(\lambda)=\mathrm{ln}\,\mathcal{Z}(\lambda)&=\mathrm{Tr}_{\omega,j}\, \mathrm{ln}\,\Big\{\mathbb{I}+4\mathbb{G}^{R}\gamma_N\mathbb{G}^{A}\gamma_1\big[\alpha_{1}\beta_{N}(e^{i\lambda}-1)+\alpha_{N}\beta_{1}(e^{-i\lambda}-1)\big]\Big\},
\end{align}
where we have used the identity $\mathrm{ln}\,\mathrm{det}[A]=\mathrm{Tr}\, \mathrm{ln}[A]$. Restoring the continuum $\omega$ framework in the steady state, we receive, 
\begin{align}
       \Tilde{\mathcal{G}}_1(\lambda) =t_M\int_{-\infty}^{\infty} \frac{d\omega}{2\pi}\,\mathrm{Tr}_{j}\, \mathrm{ln}\,\Big\{\mathcal{I}+4G^{R}(\omega)\gamma_NG^{A}(\omega)\gamma_1\big[\alpha_{1}\beta_{N}(e^{i\lambda}-1)+\alpha_{N}\beta_{1}(e^{-i\lambda}-1)\big]\Big\},\label{eq:cgf_boundary}
    \end{align}
where the origin of $t_M$ stems from the following operation; $\mathrm{Tr}_\omega[AB]=\sum_{\omega,\omega'} A_{\omega\omega'}B_{\omega'\omega}=\int\!\!\int \frac{d\omega}{2\pi}\frac{d\omega'}{2\pi}A(\omega,\omega')B(\omega',\omega)$ and in the steady-state $A(\omega,\omega')=A(\omega)\delta(\omega-\omega')$ and $B(\omega',\omega)=B(\omega')\delta(\omega-\omega')$. Now after a change of variables in the integration, one can obtain $\mathrm{Tr}_\omega[AB]=\int\frac{d\omega}{2\pi}\int\frac{d\omega'}{2\pi}A(\omega)\delta(\omega-\omega') B(\omega)\delta(\omega'-\omega)=\delta(0)\int\frac{d\omega}{2\pi}A(\omega)B(\omega)$ where $\delta(0)$ is in the units of $t_M/2\pi$. Finally the steady-state CGF for the integrated particle current $\mathcal{I}_L$ is given by,
\begin{align}
\mathcal{G}_1(\lambda)&=\frac{\Tilde{\mathcal{G}}_1(\lambda)}{t_M}=\int_{-\infty}^{\infty} \frac{d\omega}{2\pi}\,\mathrm{Tr}_{j}\, \mathrm{ln}\,\Big\{\mathcal{I}+\mathcal{T}_{1N}(\omega)\big[\alpha_{1}\beta_{N}(e^{i\lambda}-1)+\alpha_{N}\beta_{1}(e^{-i\lambda}-1)\big]\Big\},\nonumber\\
    &=\int_{-\infty}^{\infty} \frac{d\omega}{2\pi}\,{\rm ln}\,{\rm det}\,\Big\{\mathcal{I}+\mathcal{T}_{1N}(\omega)\big[\alpha_{1}\beta_{N}(e^{i\lambda}-1)+\alpha_{N}\beta_{1}(e^{-i\lambda}-1)\big]\Big\},\label{Levitov-Lesovik1}
\end{align}
where $\mathcal{I}$ is an identity matrix with dimension $N\times N$ and $\mathcal{T}_{1N}(\omega)=G^{R}(\omega)\gamma_NG^{A}(\omega)\gamma_1$ is a $N\times N$ matrix with nonzero elements only along the first column, i.e., $[\mathcal{T}_{1N}(\omega)]_{ij}=G^{R}_{iN}(\omega)\,G^{A}_{N1}(\omega)\,\delta_{j1}$. Exploiting such structure of the matrix $\mathcal{T}_{1N}(\omega)$, we can express the CGF as,
\begin{align}
    \mathcal{G}_1(\lambda)=\int_{-\infty}^{\infty} \frac{d\omega}{2\pi}\,{\rm ln}\,\Big\{1+{T}_{1N}(\omega)\big[\alpha_{1}\beta_{N}(e^{i\lambda}-1)+\alpha_{N}\beta_{1}(e^{-i\lambda}-1)\big]\Big\},\label{Levitov-Lesovik}
\end{align}
where we have used the fact that ${\rm det}\big\{\mathcal{I}+\mathcal{T}_{1N}(\omega)f(\lambda)\big\}=1+ T_{1N}(\omega)f(\lambda)$, with $T_{1N}(\omega)=4|G^{R}_{1N}(\omega)|^{2}$ and is known as the transmission function.
The above expression of the CGF resembles the structure of the well-known 
Levitov-Lesovik formula~\cite{1993JETPL..58..230L,Lesovik_2011,10.1063/1.531672} often derived  
for ideal fermionic reservoirs~\cite{PhysRevLett.90.206801,Nazarov2003,10.1063/1.5084949,Bastianello_2018,Bernard_2012,bijay2015,bijay2013}. 
Note that, following a similar procedure, as done in this section, one can also look at the statistics of the right particle current $\mathcal{I}_R$, which is the current coming into the system from $N-$th site. This requires the introduction of counting field $\lambda$ in Eq.~\eqref{Lindblad_eq} corresponding to the jump term of the dissipator $\mathcal{D}_{N}[\rho]$ as, 
\begin{align}
    \mathcal{D}^{\lambda}_{N}[\rho]=\alpha_{N}\big[2e^{i\lambda} c^{\dagger}_{N}\, \rho \,  c_{N} & -\{c_{N}c^{\dagger}_{N},\rho\} \big]+\beta_{N}\big[2e^{-i\lambda}c_{N}\rho c^{\dagger}_{N}-\{c^{\dagger}_{N}c_{N},\rho\} \big].
\end{align}
The Levitov-Lesovik formula for such a case becomes,
\begin{align}
\mathcal{G}_N(\lambda)=\int_{-\infty}^{\infty} \frac{d\omega}{2\pi}\,{\rm ln}\,\Big\{1+{T}_{N1}(\omega)\big[\alpha_{1}\beta_{N}(e^{-i\lambda}-1)+\alpha_{N}\beta_{1}(e^{i\lambda}-1)\big]\Big\},
     \label{eq:CGF_R}
\end{align}
\end{widetext}
where $T_{N1}(\omega)=4|G^{R}_{N1}(\omega)|^2$. For the setup considered in this section, $T_{N1}(\omega)=T_{1N}(\omega)$ and hence the CGF and therefore the statistics of the left current $\mathcal{I}_L$ in Eq.~\eqref{Levitov-Lesovik} and the right current $\mathcal{I}_R$ in Eq.~\eqref{eq:CGF_R} is exactly same, for arbitrary rates. This happens due to conservation of integrated current $\mathcal{N}_L$ and $\mathcal{N}_R$ across the system i.e., $\mathcal{N}_L=-\mathcal{N}_R$. Hence we will only focus on the statistics of $\mathcal{I}_L$ in this case.
However, this will not be the case when the system is subjected to particle gain and loss channels in the bulk, where the statistics of the left and the right current are distinctly different but can become identical under a set of conditions involving the rates. This we will discuss in the next section. The CGF obtained in Eq.~\eqref{Levitov-Lesovik} respects the following steady-state fluctuation symmetry \cite{Massi_RevModPhys.81.1665,Hangi2011},
\begin{align}
\mathcal{G}_1(\lambda)=\mathcal{G}_1\bigg(-\lambda+i\,\mathrm{ln}\Big\{\frac{\alpha_N\beta_1}{\beta_N\alpha_1}\Big\}\bigg).
\end{align}

\vspace{0.2cm}
\noindent {\it Verification with the known result --} We now discuss a special case of a single quantum dot coupled to two Lindblad reservoirs and obtain the expression of the CGF for the integrated current following our approach. For this simple setup, the CGF was obtained in Ref.~\cite{Massi_RevModPhys.81.1665} following an approach that relied on the eigenvalue analysis of the Liouville superoperator.

For a single quantum dot with Hamiltonian $H=\epsilon_{d}\, c^{\dagger}c$, the evolution under the GKSL master equation is given as,
 \begin{align}
     \frac{\partial \rho}{\partial t}=\mathcal{L}\rho=-i\big[\epsilon_d c^{\dagger}c,\rho&\big]+(\alpha_{1}+\alpha_{N})\big[2 c^{\dagger}\rho c-\{cc^{\dagger},\rho\} \big]\nonumber\\&\,\,\,+(\beta_{1}+\beta_{N})\big[2 c\rho c^{\dagger}-\{c^{\dagger}c,\rho\} \big].
 \end{align}
For this setup, following our analysis,  we identify the retarded and advanced Green's function for the single dot as,
\begin{align}
    G^{R}(\omega)=\frac{1}{\omega-\epsilon_{d}+i(\alpha+\beta)}=\big[G^{A}(\omega)\big]^{*},
\end{align}
where $(\alpha+\beta)$ is the sum of all the rates involved, i.e., $\alpha+ \beta=\alpha_1+\alpha_N+\beta_1+\beta_N$. To obtain a compact expression for the CGF, we work with the generalized current $\mathcal{J}_{L}(\lambda)$, given as
\begin{align}
    \mathcal{J}_{L}(\lambda)=\frac{\partial\mathcal{G}_1(\lambda)}{\partial (i\lambda)}=\int_{-\infty}^{\infty} \frac{d\omega}{2\pi} \frac{\mathcal{R'}(\lambda)}{(\omega-\epsilon_d)^{2}+\mathcal{R}(\lambda)}, \label{single_site_generalized_current}
\end{align}
where $\mathcal{R}(\lambda)=4(\alpha+\beta)^{2}+\big[\alpha_{1}\beta_{N}(e^{i\lambda}-1)+\alpha_{N}\beta_{1}(e^{-i\lambda}-1)\big]$ and $\mathcal{R'}(\lambda)=\big[\alpha_{1}\beta_{N}\,e^{i\lambda}-\alpha_{N}\beta_{1}\,e^{-i\lambda}\big]$. By employing the residue theorem, we solve the integral in Eq.~\eqref{single_site_generalized_current} exactly, that has poles at $\omega=\epsilon_{d}\pm\sqrt{\mathcal{R}(\lambda)}$ and we obtain,
\begin{align}
    \mathcal{J}_{L}(\lambda)=\frac{1}{2}\frac{\partial}{\partial(i\lambda)}\sqrt{\mathcal{R}(\lambda)}. \label{single_site_GC2}
\end{align}
Finally, the CGF is obtained as,
\begin{align}
    \mathcal{G}_1(\lambda)&=i\int_{0}^{\lambda} d\lambda \,\mathcal{J}_{L}(\lambda)=i\int_{0}^{\lambda}\!\! d\lambda \,\,\frac{1}{2}\,\frac{\partial}{\partial(i\lambda)}\sqrt{\mathcal{R}(\lambda)}\nonumber\\
    &=\frac{1}{2}\Big[\sqrt{\mathcal{R}(\lambda)}-2(\alpha+\beta)\Big].\label{cgf_singlesite}
\end{align}
The expression in Eq.~\eqref{cgf_singlesite} matches exactly with the CGF obtained in Ref.~\cite{Massi_RevModPhys.81.1665}.

\begin{figure}
    \centering   \includegraphics[width=0.5\columnwidth]{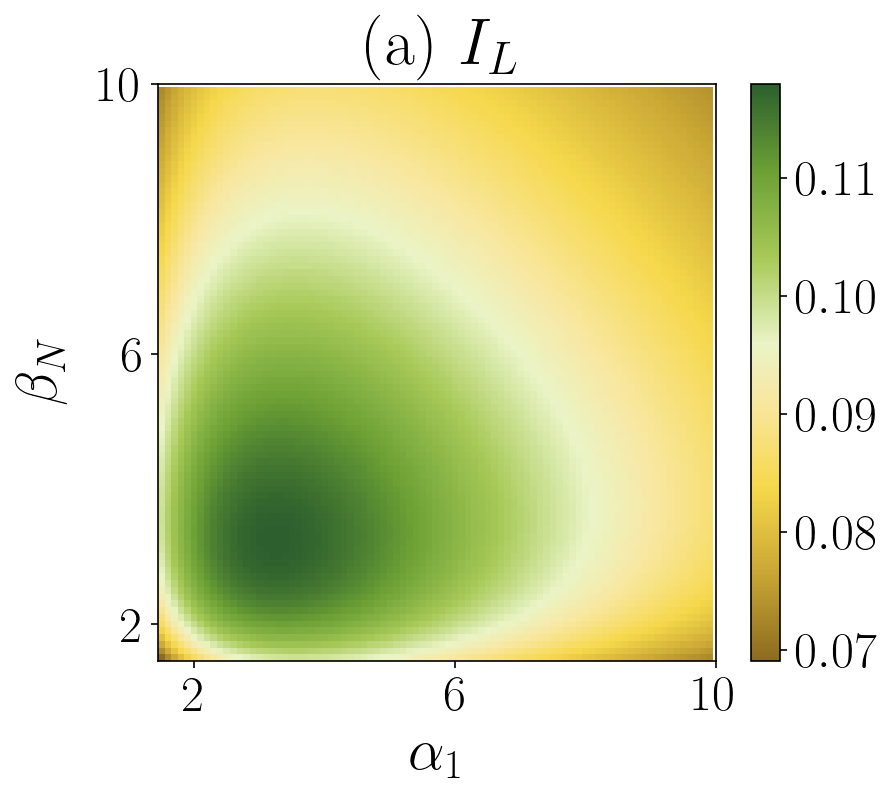}%
    \includegraphics[width=0.5\columnwidth]{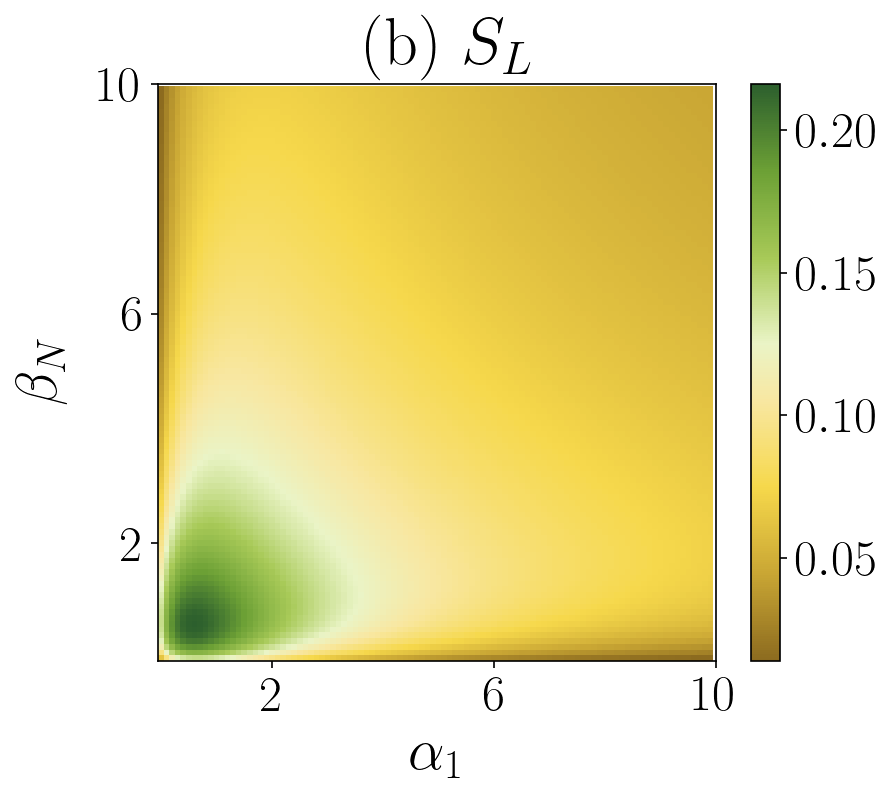}
    \caption{Colormap plot for the (a) average current $I_L$ [Eq.~\eqref{current_extended}] and (b) noise $S_L$ [Eq.~\eqref{noise_extended}] as a function of injection rate $\alpha_1$ and extraction rate $\beta_N$ for an one-dimensional tight-binding lattice. The other parameters are set as $\alpha_N=\beta_1=J=1$. The system size is taken as $N=5$. Both the current and the noise initially increases with $\alpha_1$ and $\beta_N$ and then decreases showing non-monotonic behaviour.}
    \label{S_vs_alpha}
\end{figure}
Coming back to the original expression for the CGF in Eq.~\eqref{Levitov-Lesovik}, we now write down the expressions for the first two cumulants. The first cumulant, i.e., the steady-state particle current is given by 
\begin{equation}
I_L=\langle\mathcal{I}_L\ra =\frac{\partial \mathcal{G}_1(\lambda)}{\partial (i\lambda)}\Bigg|_{\lambda=0}\!\!\!\!\!= \int_{-\infty}^{\infty}\frac{d\omega}{2\pi}\,T_{1N}(\omega)\,\big(\alpha_1 \beta_N -\beta_1 \alpha_N \big), 
\label{current_extended}
\end{equation}
where recall that $T_{1N}(\omega)=4|G^{R}_{1N}(\omega)|^{2}$ which can be interpreted as the transmission function from the left to the right end of the lattice. The factor $\big(\alpha_1 \beta_N -\beta_1 \alpha_N \big)$ serves as a bias for Lindblad type boundary drives and is analogous to the voltage bias in ideal fermionic reservoir. Positive values of this factor lead to current flow from left end to the right end of the lattice. Under reversing the values of the rates i.e., $\alpha_1 \leftrightarrow \alpha_N$ and $\beta_N \leftrightarrow \beta_1$, the bias becomes negative and as a result the directionality of the current changes to right to left without changing the magnitude of the current. This confirms that there is no specific directionality or diode like effect for the steady-state current in such non-interacting setups~\cite{Dvira2005,DviraNitzman_2005,Dvira2009}. However in the next section, we will show that there could be a situation where the current flows in reverse direction even for positive bias. For the Lindblad reservoir, the expression for steady-state current, as given in Eq.~\eqref{current_extended} was obtained recently in Ref.~\cite{Giamarchi_Lindblad}.

Next we obtain the expression of the second cumulant i.e., noise associated with the particle current which is given as,
\begin{align}
S_L=\frac{\partial^{2} \mathcal{G}_1(\lambda)}{\partial (i\lambda)^{2}}\Bigg|_{\lambda=0}&=\int_{-\infty}^{\infty}\frac{d\omega}{2\pi} \Big[T_{1N}(\omega)(\alpha_1\beta_N+\alpha_N\beta_1)\nonumber\\
&\hspace{-1em} -T^{2}_{1N}(\omega)(\alpha_1\beta_N-\alpha_N\beta_1)^{2}\Big]. \label{noise_extended}
\end{align}
We now present results for the steady-state current and corresponding noise,  obtained from Eq.~\eqref{current_extended} and Eq.~\eqref{noise_extended}, respectively, for the tight-binding model with nearest neighbor hopping in the presence of Lindblad type boundary drive.
Fig.~\ref{S_vs_alpha}, shows colormap plots of the current [Fig.~\ref{S_vs_alpha}(a)] and the noise [Fig.~\ref{S_vs_alpha}(b)] as a function of $\alpha_1$ and  $\beta_N$, where recall that $\alpha_1$ is the rate of injection in the first site and $\beta_N$ is the rate of extraction from the $N$-th site. The green region in the colormaps represent higher values of the current and noise, while the yellow region correspond to the lower values. As seen from the plots, both the current and the noise initially increases with $\alpha_1$ and $\beta_N$, reaches a maximum, and then decreases, indicating a clear
non-monotonic dependence on the boundary driving rates.
Another observation that follows from Fig.~\ref{S_vs_alpha} is that although the current remains high over a wider range of $\alpha_1$ and $\beta_N$, the corresponding noise decreases rapidly with increasing values of boundary driving rates. This suggests that it is possible to achieve large current accompanied by suppressed fluctuations. For Fig.~\ref{S_vs_alpha}, we have set other parameters $\alpha_N=1$ and $\beta_1=1$.  The current and the noise, however, display similar nonmonotonic behaviour when varied with respect to $\alpha_N$ and $\beta_1$ as well. 

Having analyzed the current and noise in presence of only boundary drive, we next add correlated gain and loss channels into the system. Since the inclusion of such channels still leads to a quadratic Liouvillian superoperator, the prescription described in this section, can be extended to such cases as well, which we will discuss in the following section.

\section{Case 2: Boundary drives + gain and loss channels in the bulk}
\label{sec:correlated}
We now adapt the previously developed formalism for a more general setup described in Eq.~\eqref{Lindblad_eq}, where the system  along with the two boundary Lindblad reservoirs, is further subjected to particle gain and loss channels in the bulk. Recall that the setup is schematically represented in Fig.~\ref{setup}. 
We follow the same procedure as described in Sec.~\ref{sec:boundary} and express the MGF $\mathcal{Z}(\lambda)$ in terms of a Path integral by dressing the dissipator $\mathcal{D}_1[\rho]$ with a counting field $\lambda$. First, the Keldysh action is obtained as [see Appendix.~\ref{sec:app2} for the derivation]
\begin{align}
\label{eq:action_gainloss}
S(\lambda)=\int_{t_0}^{t_f} dt \big[\Psi^{*}_{+}\, \big( i\,\partial_t\big) &\Psi_{+}-\Psi^{*}_{-} \, \big( i\, \partial_t \big) \, \Psi_{-}\nonumber\\&-if_\lambda(\Psi^{*}_{+},\Psi_{+},\Psi^{*}_{-},\Psi_{
    -})\big],
\end{align}
where $f_\lambda(\Psi^{*}_{+},\Psi_{+},\Psi^{*}_{-},\Psi_{-})$ is the contribution of the Liouvillian $\mathcal{L}_\lambda$ defined in Eq.~\eqref{Lindblad_eq_counting_field}. It can be decomposed in four different terms:  the Hamiltonian contribution $f_{H}(\Psi^{*}_{+},\Psi_{+},\Psi^{*}_{-},\Psi_{-})$ which is given in Eq.~\eqref{fH}, the contributions of left and right boundary drives $f_{1,N}(\Psi^{*}_{+},\Psi_{+},\Psi^{*}_{-},\Psi_{-})$ which are given in Eq.~\eqref{f1} and~\eqref{fN}.  
The fourth contribution $f_{\rm bulk}(\Psi^{*}_{+},\Psi_{+},\Psi^{*}_{-},\Psi_{-})$ is coming from the bulk dissipators i.e., from the correlated gain-loss channels. It is obtained as,
\begin{widetext}
\begin{align}
    \label{fbulk}
    f_{\rm bulk}(\Psi_{+}^{*},&\Psi_{+},\Psi_{-}^{*},\Psi_{-})=\sum_{i=1}^{N-1}\Bigg\{\Big(\Gamma_1\big[2\psi_{+}^{i*}\psi^{i}_{-}+\psi_{+}^{i*}\psi_{+}^{i}+\psi_{-}^{i*}\psi_{-}^{i}\big]+\Gamma_2\big[2\psi_{+}^{i+1*}\psi_{-}^{i+1}+\psi_{+}^{i+1*}\psi_{+}^{i+1}+\psi_{-}^{i+1*}\psi_{-}^{i+1}\big]\nonumber\\&+e^{-i\theta}\sqrt{\Gamma_1\Gamma_2}\big[2\psi_{+}^{i*}\psi_{-}^{i+1}+\psi_{+}^{i+1*}\psi_{+}^{i}+\psi_{-}^{i+1*}\psi_{-}^{i}\big]+e^{i\theta}\sqrt{\Gamma_1\Gamma_2}\big[2\psi_{+}^{i+1*}\psi_{-}^{i}+\psi_{+}^{i*}\psi_{+}^{i+1}+\psi_{-}^{i*}\psi_{-}^{i+1}\big]\Big)\nonumber\\&+\Big(\kappa_1\big[2\psi_{+}^{i}\psi_{-}^{i*}-\psi_{+}^{i*}\psi_{+}^{i}-\psi_{-}^{i*}\psi_{-}^{i}\big]+\kappa_2\big[2\psi_{+}^{i+1}\psi_{-}^{i+1*}-\psi_{+}^{i+1*}\psi_{+}^{i+1}-\psi_{-}^{i+1*}\psi_{-}^{i+1}\big]\nonumber\\&+e^{-i\phi}\sqrt{\kappa_1\kappa_2}\big[2\psi_{+}^{i}\psi_{-}^{i+1*}-\psi_{+}^{i+1*}\psi_{+}^{i}-\psi_{-}^{i+1*}\psi_{-}^{i}\big]+e^{i\phi}\sqrt{\kappa_1\kappa_2}\big[2\psi_{+}^{i+1}\psi_{-}^{i*}-\psi_{+}^{i*}\psi_{+}^{i+1}-\psi_{-}^{i*}\psi_{-}^{i+1}\big]\Big)\Bigg\}.
\end{align}
We then follow the procedure described in Sec.~\ref{sec:boundary} in Eq.~\eqref{action_matrix_form} -- Eq.~\eqref{eq:mlambda} and identify various Green's function as,
\begin{align}
    G^{R}(\omega)=\Big[\omega \mathcal{I}-h+i(\gamma^{R}_{1}+\gamma^{R}_{N}+\gamma^{R}_{\rm bulk})\Big]^{-1}= \Big[G^{A}(\omega)\Big]^{\dagger},\,\, {\rm and}\,\, G^{K}(\omega)=\, i\, G^{R}(\omega)\,\big[\gamma^{K}_{1}+\gamma^{K}_{N}+\gamma^{K}_{\rm bulk}\big]\,G^{A}(\omega), \label{eq:gr_gk}
\end{align}
where $\mathcal{I}$ is the identity matrix of dimension $N\times N$.
$\gamma^{R}_{1,N}$ and $\gamma^{K}_{1,N}$ are the self energies corresponding to the left and right reservoirs and they are same as given in Eq.~\eqref{eq:ret_selfenergy} and Eq.~\eqref{eq:Keldysh_selfenergy}, respectively. The self energy components corresponding to the correlated gain-loss channels are given by,
\begin{align}
    &\gamma^{R}_{{\rm bulk},ij}=\Big[(\Gamma_1+\kappa_1+\Gamma_2+\kappa_2)\delta_{ij}+(e^{i\theta}\sqrt{\Gamma_1\Gamma_2} + e^{-i\phi}\sqrt{\kappa_1\kappa_2} )\delta_{i,j-1}+(e^{-i\theta}\sqrt{\Gamma_1\Gamma_2}+ e^{i\phi}\sqrt{\kappa_1\kappa_2})\delta_{i,j+1}\Big],\label{eq:self_energy_R_bulk} \\
    &\gamma^{K}_{{\rm bulk},ij}=2\Big[(\Gamma_1-\kappa_1+\Gamma_2-\kappa_2)\delta_{ij}+(e^{i\theta}\sqrt{\Gamma_1\Gamma_2} - e^{-i\phi}\sqrt{\kappa_1\kappa_2} )\delta_{i,j-1}+(e^{-i\theta}\sqrt{\Gamma_1\Gamma_2}- e^{i\phi}\sqrt{\kappa_1\kappa_2})\delta_{i,j+1}\Big].\label{eq:self_energy_K_bulk}
\end{align}
Note that these bulk self energies are tridiagonal matrices and in particular the off diagonal elements are solely due to the presence of correlated dissipators. For the case of localized gain-loss channels ($\Gamma_2=0$ and $\kappa_2=0$), the self energies are diagonal i.e, $\gamma_{{\rm bulk},ij}^{R,K}=(\kappa_1\pm\Gamma_1)\delta_{ij}$. The presence of the off-diagonal elements in the self energies has a profound implication in the steady-state current statistics which we will discuss later.
Finally, once identified the Green's functions, 
we 
can obtain the CGF corresponding to the current $\mathcal{I}_L$ and $\mathcal{I}_R$ for this setup, following the procedure in the previous section. The CGF corresponding to the current $\mathcal{I}_L$ is obtained as,
\begin{equation}
\!\!\mathcal{G}_1(\lambda)\!=\!\int_{-\infty}^{\infty} \!\frac{d\omega}{2\pi}\, \mathrm{ln}\,\Big\{1+ {T}_{1N}(\omega)\,\big[\alpha_{1}\beta_{N}(e^{i\lambda}-1)+\alpha_{N}\beta_{1}(e^{-i\lambda}-1)\big]\!+\!\big[{T}^{>}_{1}(\omega)\alpha_{1}(e^{i\lambda}-1)+{T}^{<}_{1}(\omega)\beta_{1}(e^{-i\lambda}-1)\big]\Big\}.\label{CGF_gain_lossL}
\end{equation}
In a similar way, the CGF corresponding to the particle current $\mathcal{I}_R$ can be obtained as,
\begin{equation}
\!\!\mathcal{G}_N(\lambda)\!=\!\int_{-\infty}^{\infty}\! \frac{d\omega}{2\pi} \, \mathrm{ln}\,\Big\{1+{T}_{N1}(\omega)\big[\alpha_{1}\beta_{N}(e^{-i\lambda}-1)+\alpha_{N}\beta_{1}(e^{i\lambda}-1)\big]\!\!+\!\big[{T}^{>}_{N}(\omega)\alpha_{N}(e^{i\lambda}-1)+{T}^{<}_{N}(\omega)\beta_{N}(e^{-i\lambda}-1)\big]\Big\}.\label{CGF_gain_lossR}
\end{equation}
\end{widetext}
Here, ${T}_{1N}(\omega)=4\, \mathrm{Tr}\big[G^{R}(\omega){\gamma}_NG^{A}(\omega){\gamma}_1\big]$, and ${T}_{N1}(\omega)=4\, \mathrm{Tr}\big[G^{R}(\omega){\gamma}_1G^{A}(\omega){\gamma}_N\big]$ are the transmission functions from $1$st to $N-$th site and $N-$th to $1$st site, respectively. Their expressions can be simplified and written as,
\begin{equation}
T_{1N}(\omega)=4|G^{R}_{1N}(\omega)|^{2}, \quad T_{N1}(\omega)=4|G^{R}_{N1}(\omega)|^2. \label{eq:transmission_gainloss}
\end{equation} 
The terms ${T}^{>}_{1,(N)}(\omega)=-2\mathrm{Tr}\big[G^{R}(\omega)\gamma_{\rm bulk}^{>}G^{A}(\omega)\gamma_{1,(N)}\big]$ and ${T}_{1,(N)}^{<}(\omega)=2\mathrm{Tr}\big[G^{R}(\omega)\gamma_{\rm bulk}^{<}G^{A}(\omega)\gamma_{1,(N)}\big]$ describe the transmission functions from the $1$st ($N$-th) site to the loss channels and the gain channels, respectively. Here we call $\gamma_{\rm bulk}^{>,<}$ as the greater and lesser self energy components due to the correlated dissipators and they are given by,
\begin{align}
    \gamma_{{\rm bulk},lm}^{>}=-2\big[(\kappa_1+\kappa_2)\, \delta_{lm}&+e^{-i\phi}\sqrt{\kappa_1\kappa_2}\,\delta_{l(m-1)}\nonumber\\& \hspace{-2em} +e^{i\phi}\sqrt{\kappa_1\kappa_2}\,\delta_{l(m+1)}\big],\label{self_energy_gre_bulk}\\
    \gamma_{{\rm bulk},lm}^{<}=2\big[(\Gamma_1+\Gamma_2)\,\delta_{lm}&+e^{i\theta}\sqrt{\Gamma_1
    \Gamma_2}\,\delta_{l(m-1)}\nonumber\\&
     \hspace{-3em} +e^{-i\theta}\sqrt{\Gamma_1\Gamma_2}\,\delta_{l(m+1)}\big].\label{self_energy_les_bulk}
\end{align}
The CGFs in Eq.~\eqref{CGF_gain_lossL} and \eqref{CGF_gain_lossR} encapsulate the current statistics in the presence of gain-loss channels.
It is worth mentioning that these CGFs of the currents $\mathcal{I}_L$ and $\mathcal{I}_R$ are in general not same because of the presence of ${T}_{1,N}^{<,>}(\omega)$ terms. Furthermore, the transmission functions $T_{1N}(\omega)$, and $T_{N1}(\omega)$, in general, differ in presence of gain-loss channels, which further contributes to the asymmetry in the statistics of $\mathcal{I}_L$ and $\mathcal{I}_R$. 

From the two CGFs in Eq.~\eqref{CGF_gain_lossL} and~\eqref{CGF_gain_lossR}, we can compute all the  cumulants. The first cumulant, i.e., the average left and right currents $I_L$, $I_R$ are given as 
\begin{align}
    I_L = \frac{\partial\mathcal{G}_1(\lambda)}{\partial(i\lambda)}\Bigg|_{\lambda=0}\!\!\!\!\!=\int_{-\infty}^{\infty}\frac{d\omega}{2\pi} &\Big[T_{1N}(\omega)(\alpha_1\beta_N -\alpha_N\beta_1)\nonumber\\
    &\hspace{-3em}+T_{1}^{>}(\omega)\alpha_1-T_{1}^{<}(\omega)\beta_1\Big], \label{eq:I_L_expr}\\
 I_R = \frac{\partial\mathcal{G}_N(\lambda)}{\partial(i\lambda)}\Bigg|_{\lambda=0}\!\!\!\!\!=\int_{-\infty}^{\infty}\frac{d\omega}{2\pi} &\Big[T_{N1}(\omega)(\alpha_N\beta_1 -\alpha_1\beta_N)\nonumber\\
    & \hspace{-3em}+T_{N}^{>}(\omega)\alpha_N-T_{N}^{<}(\omega)\beta_N\Big].\label{eq:I_R_expr}
\end{align}
Here we follow the convention that current coming out from any of the end is considered to be positive. One can further construct the net current that flows from left end to the right end of the lattice. This is given as $I_{LR}=\big(I_L - I_R\big)/2$. Explicity, this can be written down as 
\begin{align}
    I_{LR}&=\int_{-\infty}^{\infty}\frac{d\omega}{2\pi}\Bigg[\frac{T_{1N}(\omega)+T_{N1}(\omega)}{2}(\alpha_1\beta_N-\beta_1\alpha_N)\Bigg]\nonumber\\
    & \hspace{-2.5em} +\frac{1}{2}\Bigg[T_{1}^{>}(\omega)\,\alpha_1 \!-\! T_{1}^<(\omega)\, \beta_1 \!-\! T_{N}^{>}(\omega)\,\alpha_N \!+\! T_{N}^{<}(\omega)\,\beta_N\Bigg].\label{eq:left_to_right_current}
\end{align}
Another interesting quantity which can be constructed from $I_L$ and $I_R$ is the total bulk current $I_{\rm bulk}=I_L+I_R$ which is going out or coming into the system and it is defined as,
\begin{align}
    I_{\rm bulk}=\int_{-\infty}^{\infty}\frac{d\omega}{2\pi}& \Big[\big(T_{1N}(\omega)-T_{N1}(\omega)\big)(\alpha_1\beta_N - \alpha_N\beta_1)\nonumber\\
    &\hspace{-5.5em}+T_{1}^{>}(\omega)\alpha_1-T_{1}^<(\omega)\beta_1+T_{N}^{>}(\omega)\alpha_N-T_{N}^{<}(\omega)\beta_N\Big].\label{eq:bulk_current}
\end{align}
Note that $I_{\rm bulk}$ is always zero for the boundary driven setup in absence of any loss and gain channels in the bulk. However, in presence of such gain-loss channels, in general, it is non-zero.
The second cumulant i.e., the fluctuation of the current is obtained by taking second order derivative of the respective CGF with respect to $i\lambda$. The fluctuation of the current $\mathcal{I}_L$ is given as,
\begin{align}
    S_L\!\! =&\!\!\int_{-\infty}^{\infty}\!\frac{d\omega}{2\pi}\!\Big[{T}_{1N}(\omega)\big(\!\alpha_1\beta_N\!+\!\alpha_N\beta_1\!\big)\!+\!{T}_1^{>}(\omega)\alpha_1\!+\!{T}_1^{<}(\omega)\beta_1\!\Big]\nonumber\\
    &\hspace{-2em}-\Big[{T}_{1N}(\omega)\big(\!\alpha_1\beta_N\!-\!\alpha_N\beta_1\!\big)\!+\!{T}_{1}^{>}(\omega)\alpha_1\!-\!{T_{1}^{<}}(\omega)\beta_1\!\Big]^2\!. \label{eq:S_L}
\end{align}
Similarly, the fluctuation in the current $I_R$ is given as,
\begin{align}
    S_R\! =&\!\!\int_{-\infty}^{\infty}\!\!\frac{d\omega}{2\pi}\!\Big[\!{T}_{N1}(\omega)\big(\!\alpha_1\beta_N\!+\!\alpha_N\beta_1\!\big)\!+\!{T}_N^{>}(\omega)\alpha_N\!+\!{T}_N^{<}(\omega)\beta_N\!\Big]\nonumber\\
    &\hspace{-3em}-\Big[{T}_{N1}(\omega)\big(\!\alpha_1\beta_N\!-\!\alpha_N\beta_1\!\big)\!-\!{T}_{N}^{>}(\omega)\alpha_N\!+\!{T_{N}^{<}}(\omega)\beta_N\!\Big]^2\!.\label{eq:S_R}
\end{align}
In the following, we first analyze the statistics of $\mathcal{I}_L$ and $\mathcal{I}_R$ for the localized gain-loss case, obtained by setting $\Gamma_1=\Gamma$, $\kappa_1=\kappa$, and $\Gamma_2=\kappa_2=0$. We then examine the correlated gain-loss case by setting $\Gamma_1=\Gamma_2=\Gamma$, $\kappa_1=\kappa_2=\kappa$, without loss of generality.

\subsection{Localized gain-loss channels}\label{subsec:localloss}

\begin{figure*}
    \centering
    \includegraphics[width=0.54\columnwidth]{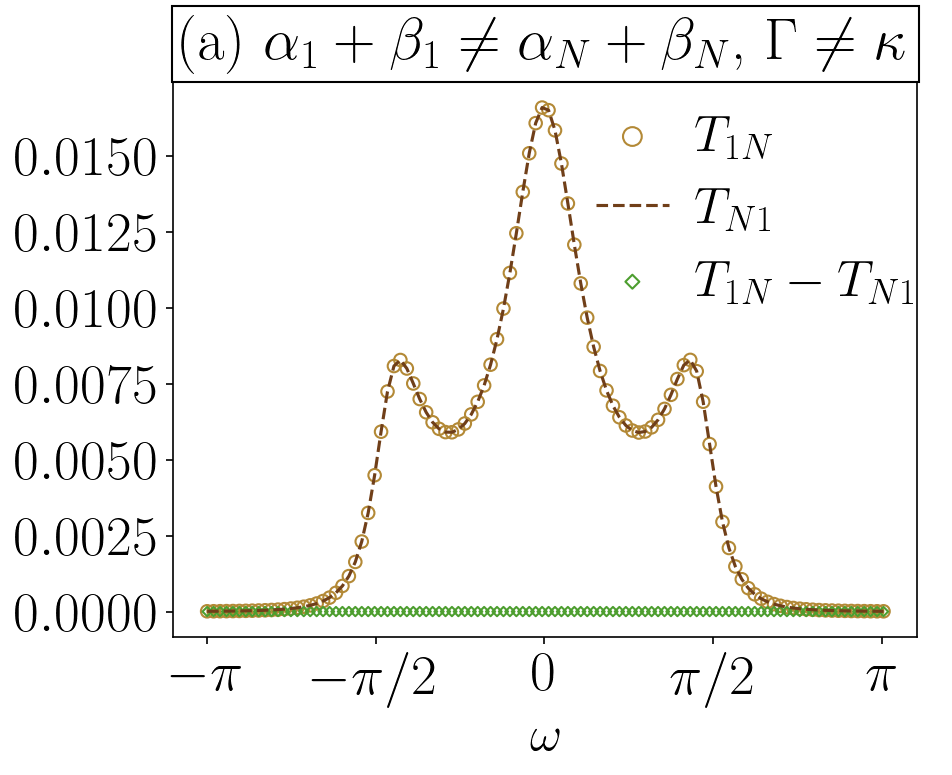}%
    \includegraphics[width=0.5\columnwidth]{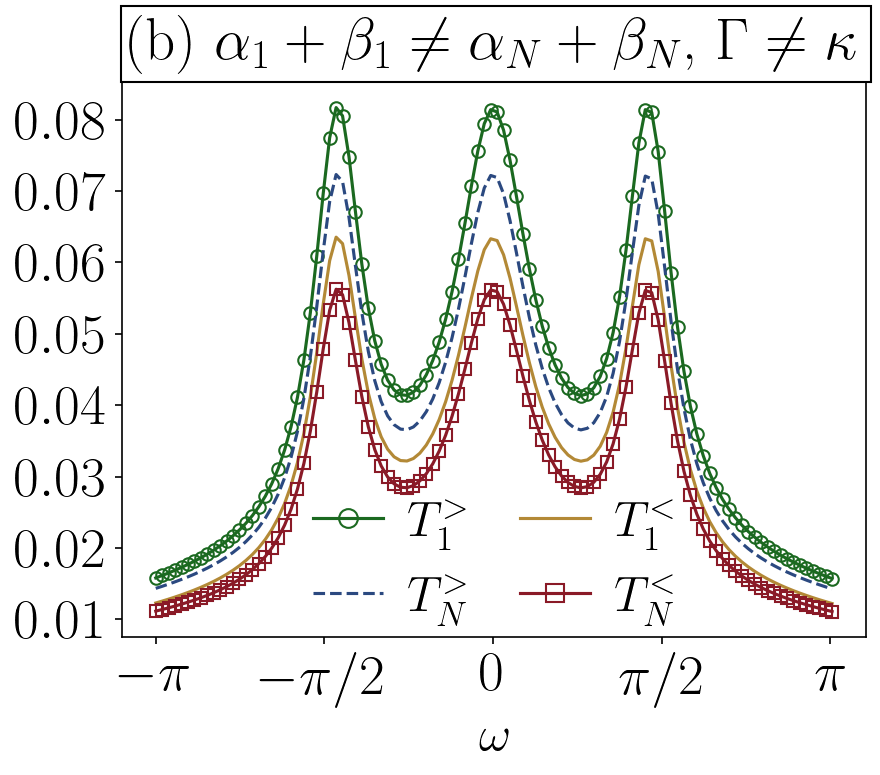}%
    \includegraphics[width=0.515\columnwidth]{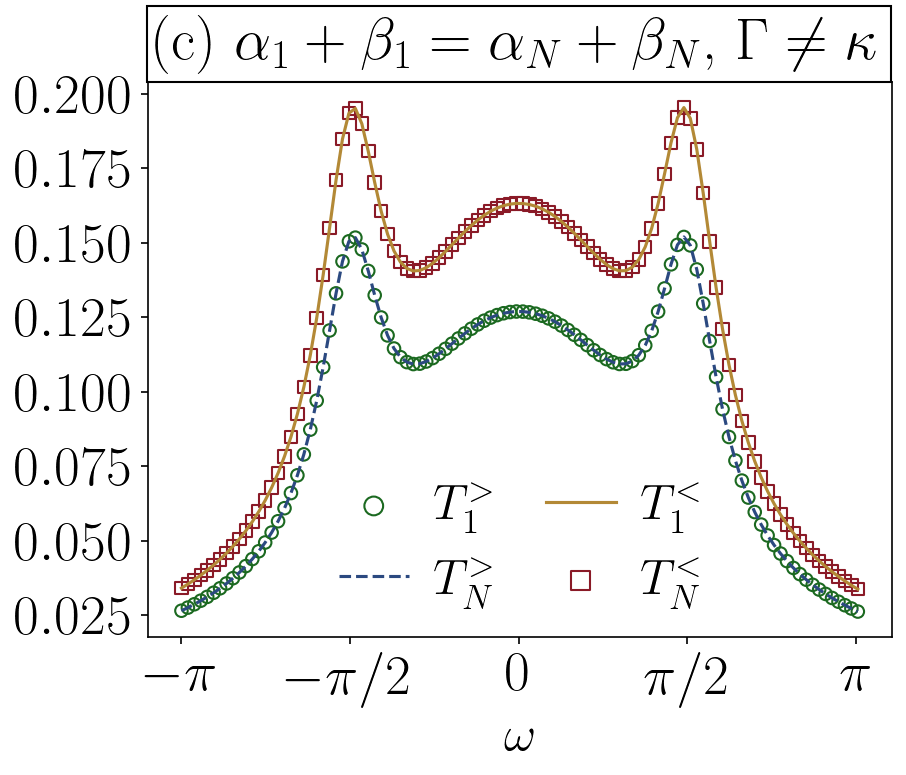}%
    \includegraphics[width=0.5\columnwidth]{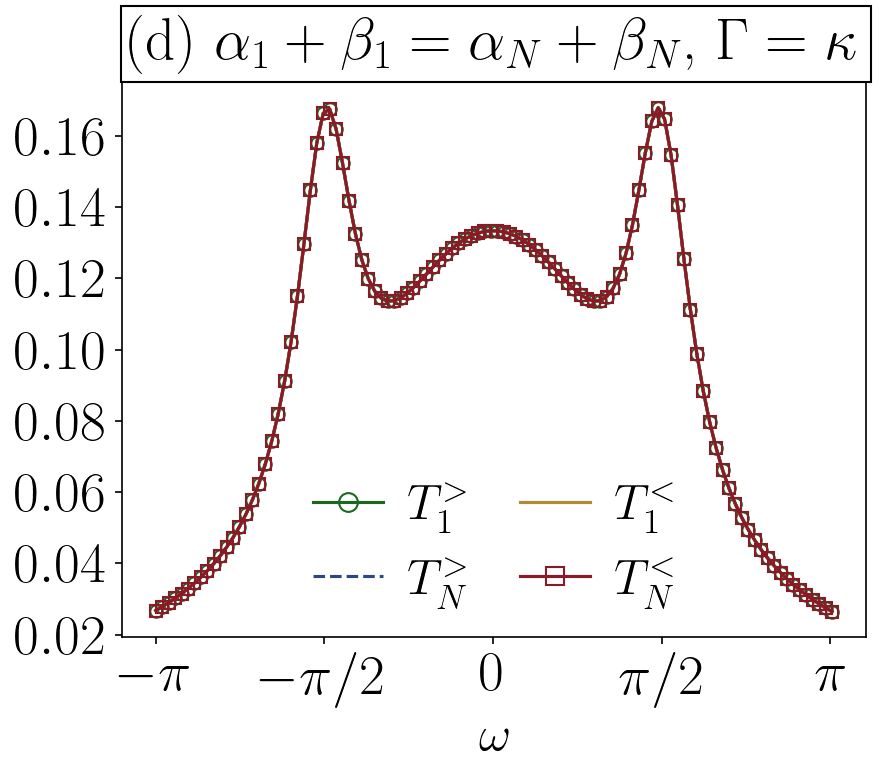}%
    \caption{Local gain-loss case: plot of different transmission functions that appears in the CGF of $\mathcal{I}_L$ and $\mathcal{I}_R$ in Eq.~\eqref{CGF_gain_lossL} and \eqref{CGF_gain_lossR}. (a) Transmission functions from the left end to the right end, $T_{1N}(\omega)$ [Eq.~\eqref{eq:transmission_gainloss}] and from the right end to the left end $T_{N1}(\omega)$ [Eq.~\eqref{eq:transmission_gainloss}] are plotted with frequency $\omega$ for the parameters $\alpha_1=4.0$, $\beta_1=0.1$, $\alpha_N=1.9$, $\beta_N=2.5$, $\Gamma=0.07$, and $\kappa=0.09$. The system size is chosen to be $N=5$. Their difference is zero for all values of $\omega$ which confirms that in presence of localized gain-loss channels, $T_{1N(\omega)}=T_{N1}(\omega)$ for any choice of $\alpha_1$, $\beta_1$, $\alpha_N$, $\beta_N$, $\Gamma$, and $\kappa$. For the same set of parameters, the transmission functions from $1$st site to the gain loss channels i.e., $T_1^{>}$ and $T_1^{<}$ [see Eq.~\eqref{eq:t1_gl}],  from $N$-th site to the gain-loss channels i.e., $T_{N}^{>}$ and $T_{N}^<$ [see Eq.~\eqref{eq:tn_gl}] are plotted in (b). It can be clearly seen that all the transmission functions are different as $\alpha_1+\beta_1$ is different from $\alpha_N+\beta_N$ and $\Gamma \neq \kappa$. (c) Once the condition $\alpha_1+\beta_1=\alpha_N+\beta_N$ is introduced by the choice of $\alpha_1=1.0=\beta_N$ and $\beta_1=\alpha_N=0.9$, one can see from this plot that $T_1^{>}=T_{N}^{>}$ and $T_{1}^<=T_{N}^<$. However as $\Gamma = 0.09$ and $\kappa=0.07$, all four of them are not equal to each other. (c) Once we further choose $\Gamma=\kappa=0.07$, all four transmission functions become equal and under these conditions the left and right current statistics become identical.}
    \label{fig:transmission_local}
\end{figure*}

In presence of localized gain-loss channels at every site of the lattice, all the bulk self energies $\gamma_{\rm bulk}^{R}$, $\gamma_{\rm bulk}^{K}$, $\gamma^{>}_{\rm bulk}$, and $\gamma_{\rm bulk}^{<}$ in Eq.~\eqref{eq:self_energy_R_bulk},~\eqref{eq:self_energy_K_bulk},~\eqref{self_energy_gre_bulk}, and~\eqref{self_energy_les_bulk} respectively, are diagonal. This leads to further simplifications in the expression of transmission functions ${T}_{1,N}^{>,<}(\omega)$. Specifically,
\begin{align}
     &T_{1}^{>,<}(\omega) = \sum_{n=1}^{N} 2\gamma^{>,<}_{nn} |G^{R}_{1n}(\omega)|^2, \label{eq:t1_gl}\\
    &T_{N}^{>,<}(\omega) = \sum_{n=1}^{N} 2\gamma^{>,<}_{nn} |G^{R}_{Nn}(\omega)|^2,\label{eq:tn_gl}
\end{align}
where $G^{R}(\omega)$, defined in Eq.~\eqref{eq:gr_gk}, is the inverse of a symmetric tridiagonal matrix for each $\omega$. Using the transfer matrix technique, as adapted in Ref.~\cite{loss2024}, one can show that, such symmetric matrices gives the relation $|G^{R}_{1N}(\omega)|=|G^{R}_{N1}(\omega)|$ which in turn implies $T_{1N}(\omega)=T_{N1}(\omega)$ from Eq.~\eqref{eq:transmission_gainloss}. In Fig.~\ref{fig:transmission_local}(a), we  show the equality between left-to-right transmission function $T_{1N}(\omega)$ and the right-to-left transmission function $T_{N1}(\omega)$ for the localized gain-loss case. 
It is not possible to alter this equality even by varying the parameters $\alpha_1$, $\beta_1$, $\alpha_N$, $\beta_N$, $\Gamma$, and $\kappa$. This implies that the reciprocity cannot be broken in presence of localized gain-loss channels. Note that, although $T_{1N}(\omega)=T_{N1}(\omega)$ in this case, the transmission functions from the $1$st and $N$-th site to the gain-loss channels i.e., $T^{<,>}_1(\omega)$ and $T_{N}^{<,>}(\omega)$ are in general unequal [see Fig.~\ref{fig:transmission_local}(b)]. As a consequence, the statistics of $\mathcal{I}_L$ and $\mathcal{I}_R$ at the two boundaries are, in general, different.
This behaviour is in stark contrast with the ``no  gain-loss'' case discussed in Sec.~\ref{sec:boundary}, where the statistics are identical for arbitrary parameter values. Interestingly, in the local gain-loss scenario, identical statistics of $\mathcal{I}_L$ and $\mathcal{I}_R$ can still be obtained by choosing appropriate parameters. In particular, when the condition $\alpha_1+\beta_1=\alpha_N+\beta_N$ is imposed, one can show using the transfer matrix approach that,
$|G^{R}_{1n}(\omega)|=|G^{R}_{N,N-n+1}(\omega)|$~\cite{loss2024}. This relation directly leads to $T_1^{>}(\omega)=T_{N}^{>}(\omega)$ and $T^{<}_1(\omega)=T_{N}^{<}(\omega)$, as illustrated in Fig.~\ref{fig:transmission_local}(c), where the transmission functions \big[$T^{<,>}_1(\omega),T_{N}^{<,>}(\omega)$\big] are plotted with the frequency $\omega$. Furthermore when the system possesses balance loss and gain in the bulk, i.e., $\kappa=\Gamma$, i.e., one obtains 
$T^{>}_1(\omega)=T^{<}_{1}(\omega)=T^{>}_N(\omega)=T^{<}_{N}(\omega)$, as clearly shown in fig.~\eqref{fig:transmission_local}(d). 
Finally, to achieve identical CGF for $\mathcal{I}_L$ in Eq.~\eqref{CGF_gain_lossL} and $\mathcal{I}_R$ in Eq.~\eqref{CGF_gain_lossR},  one must additionally impose $\alpha_1=\beta_N$ and $\beta_1=\alpha_N$. Below we summarize these three conditions required to ensure identical statistics for $\mathcal{I}_L$ and $\mathcal{I}_R$:
\begin{enumerate}[label=\roman*.,ref=(\roman*)]
    \item $\alpha_1+\beta_1=\alpha_N+\beta_N$, \quad (Reflection symmetry of the boundary couplings) \label{cond:A}
    \item $\kappa=\Gamma$, \quad (Balanced gain and loss in the bulk of the lattice) \label{cond:B}
    \item $\alpha_1=\beta_N$, $\alpha_N=\beta_1$, \quad (Balanced gain and loss at the two boundaries) \label{cond:C}
\end{enumerate}
The above three conditions results in a balanced gain-loss scenario in which the system respects the $\mathcal{PT}$ symmetry (the first condition ensure parity, while the second and third conditions correspond to balanced gain and loss) and thereby leading to identical statistics for the left and the right current. Note that in such a scenario, following Eq.~\eqref{eq:bulk_current}, $I_{\rm bulk}=0$. Under this scenario, if we interchange the boundary rates ($\alpha_1\leftrightarrow\alpha_N$ and $\beta_1\leftrightarrow\beta_N$), the magnitude of the current does not change. 
In this sense the setup is {\it reciprocal} and hence {\it no} diode effect can be observed.

\begin{figure*}
    \centering
    \includegraphics[width=5.5cm]{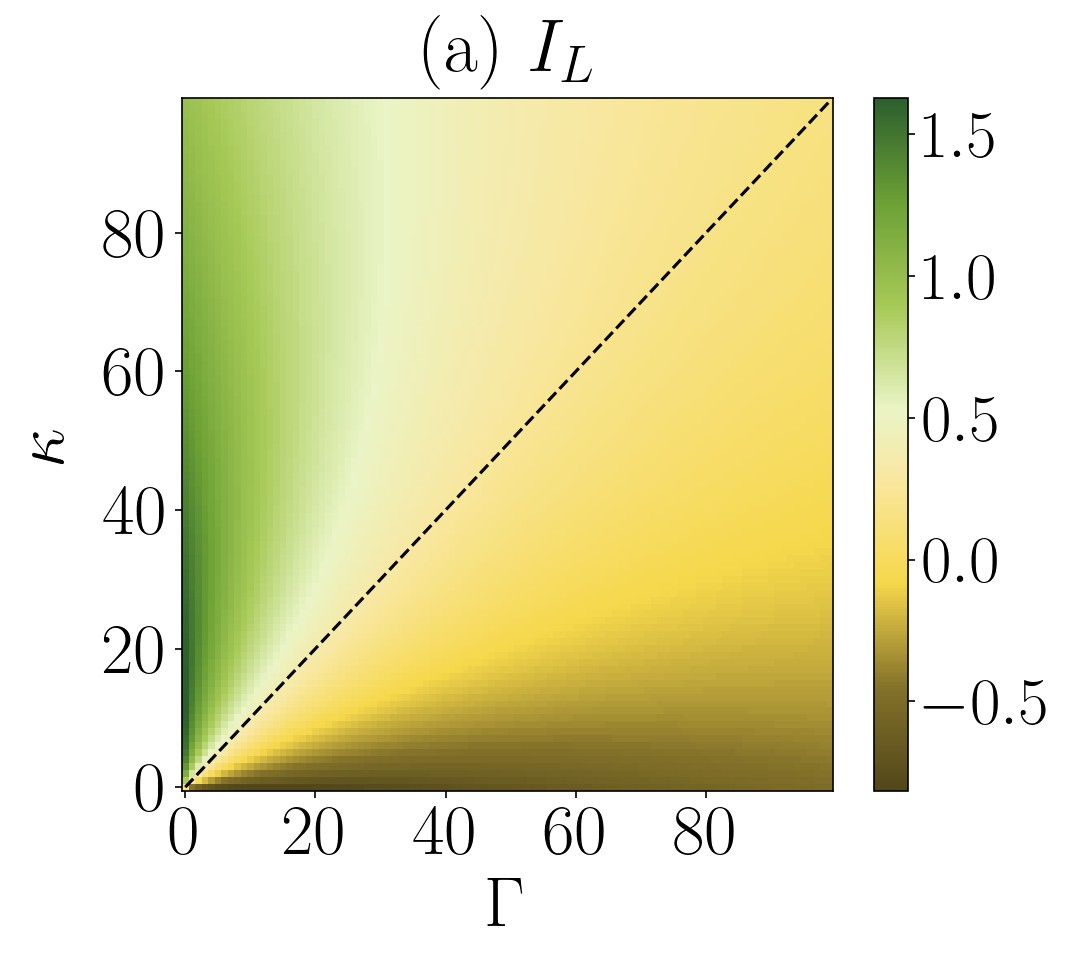}%
    \includegraphics[width=5.5cm]{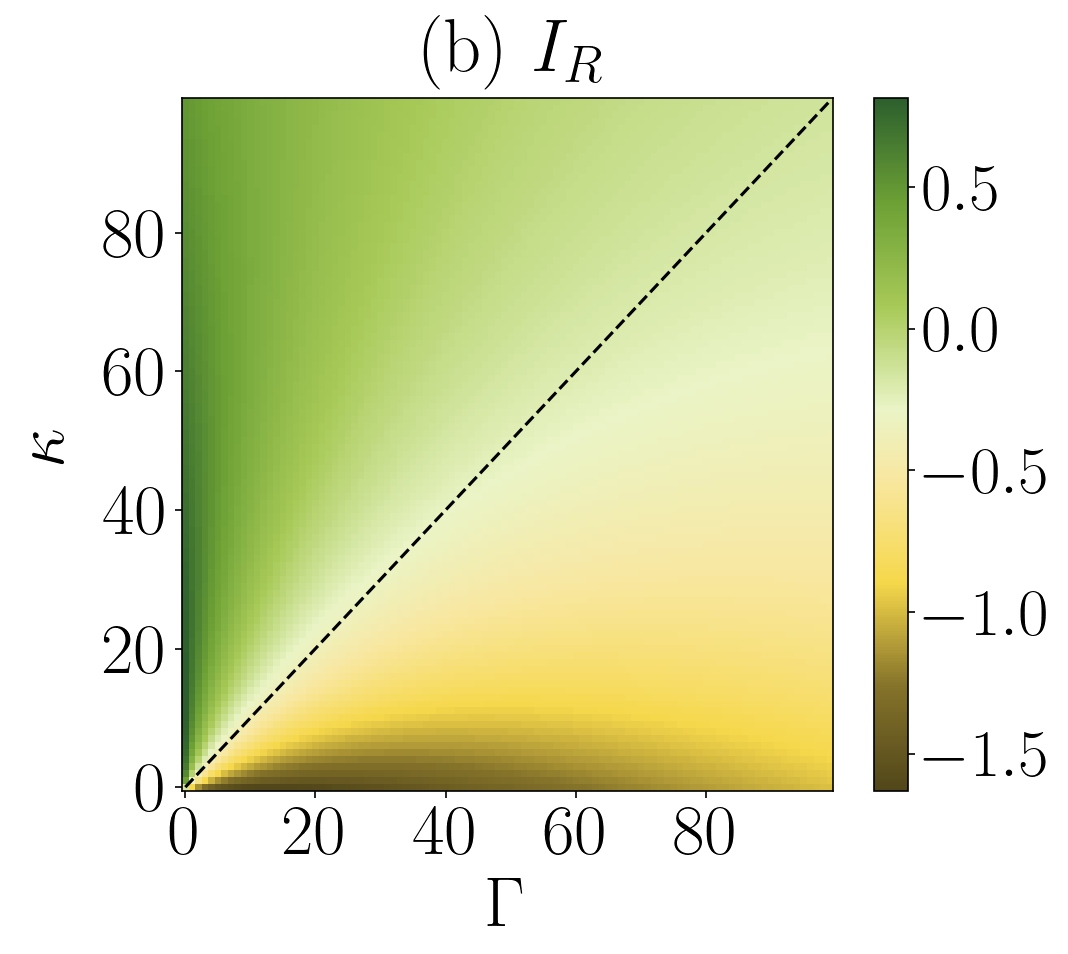}%
    \includegraphics[width=5.2cm]{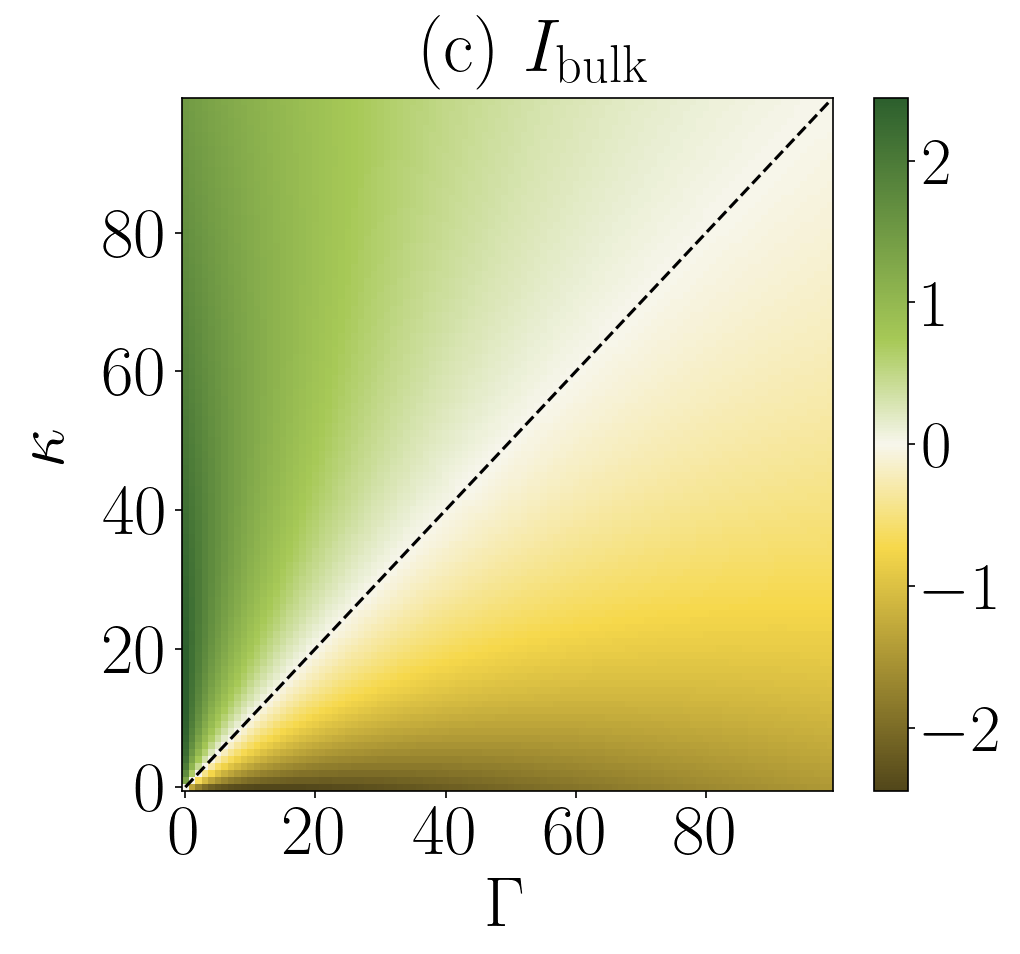}
    \includegraphics[width=5.5cm]{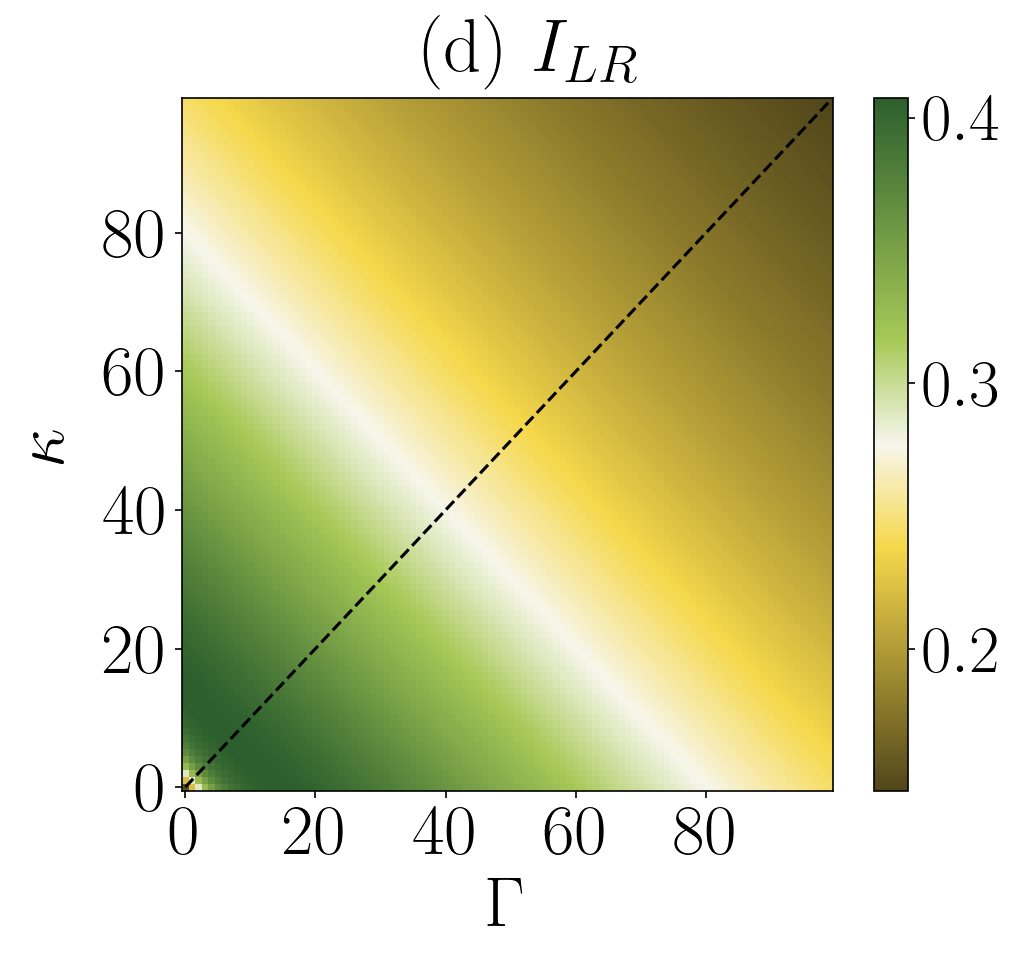}%
    \includegraphics[width=5.5cm]{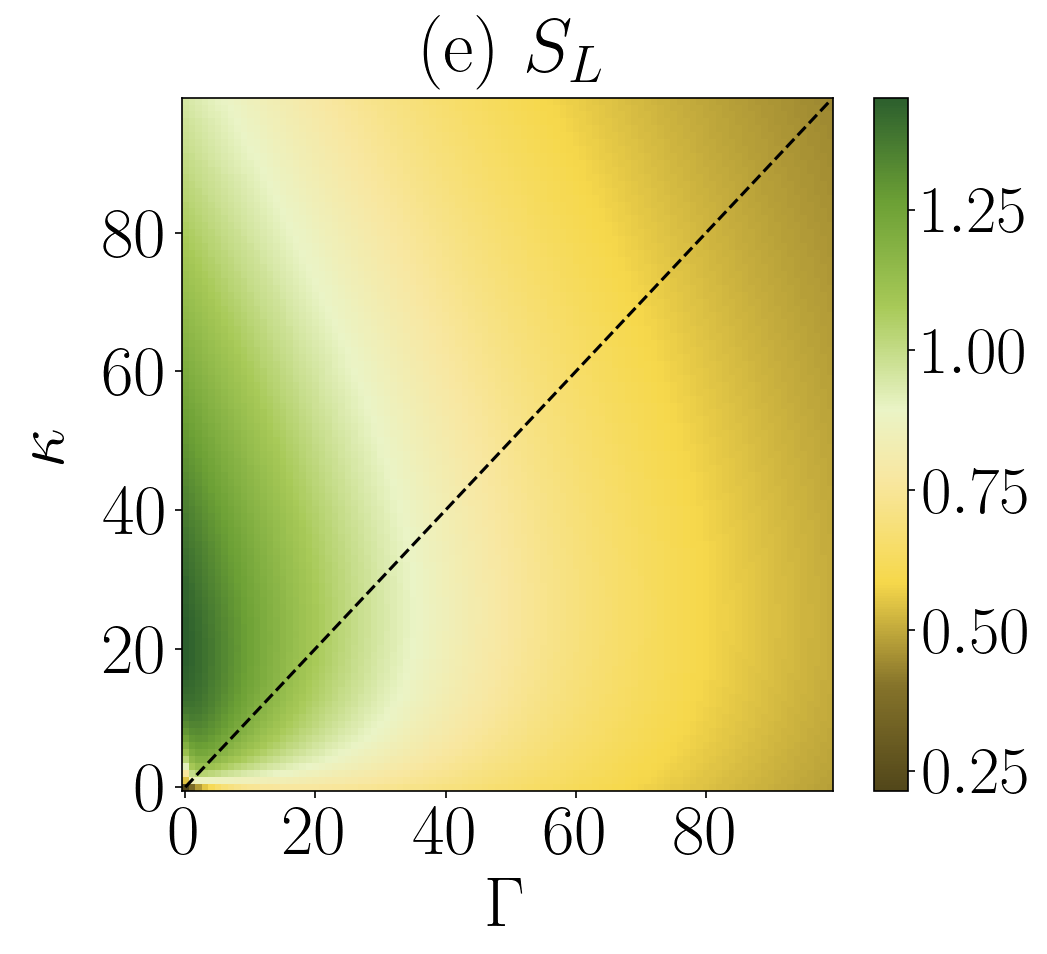}%
    \includegraphics[width=5.3cm]{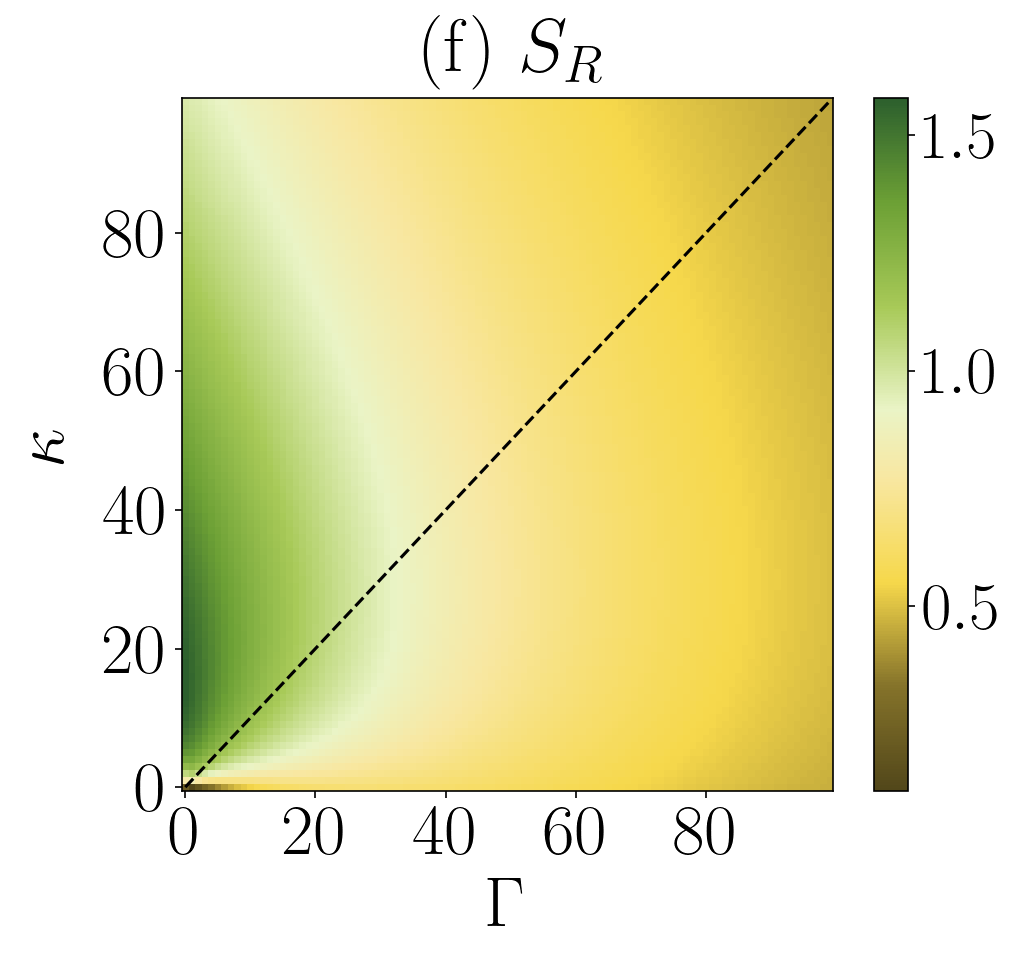}%
    \caption{Local gain-loss case: colormap plot of the statistics of left and right current in local gain-loss case [Subsec.~\ref{subsec:localloss}]. The parameters chosen here are $\alpha_1=\beta_N=1.0$ and $\alpha_N=\beta_1=0.5$ which satisfy the condition $\alpha_1+\beta_1=\alpha_N+\beta_N$. (a) Plot of average current $I_L$ using Eq.~\eqref{eq:I_L_expr}, (b) fluctuations $S_L$ using Eq.~\eqref{eq:S_L} corresponding to the left current $\mathcal{I}_L$, (c) average current $I_R$ using Eq.~\eqref{eq:I_R_expr}, (d) noise $S_R$ using Eq.~\eqref{eq:S_R} corresponding to the right current $I_R$. The statistics of $\mathcal{I}_L$ and $\mathcal{I}_R$ are in general different which can be seen from the plots. Along the line $\Gamma=\kappa$, (black dashed line) the statistics coincide. (e) Plot of the net left to right current $I_{LR}$ defined in Eq.~\eqref{eq:left_to_right_current}. There is no diode-like effect in this case. (f) Plot of the total bulk current $I_{\rm bulk}=I_L+I_R$ which is going out or coming into the system and it is defined in Eq.~\eqref{eq:bulk_current}. $I_{\rm bulk}$ is zero along the $\Gamma=\kappa$ line (black dashed line) which is the balanced gain-loss scenario.}
    \label{fig:I_L_stat}
\end{figure*}

We now use the expressions of average left current $I_L$ [Eq.~\eqref{eq:I_L_expr}], average right current $I_R$ [Eq.~\eqref{eq:I_R_expr}] and their respective fluctuations $S_L$ [Eq.~\eqref{eq:S_L}] and $S_R$ [Eq.~\eqref{eq:S_R}], and present the corresponding results in Fig.~\ref{fig:I_L_stat} for this case. 
Here we enforce only conditions~\ref{cond:A} and~\ref{cond:C} in our analysis by choosing the boundary rates as $\alpha_1=\beta_N=1.0$ and $\alpha_N=\beta_1=0.5$ while varying $\kappa$ and $\Gamma$. We illustrate the statistics of $\mathcal{I}_L$ and $\mathcal{I}_R$ in Fig.~\ref{fig:I_L_stat}: the average left current $I_L$ is shown in Fig.~\ref{fig:I_L_stat}(a) and its corresponding fluctuations $S_L$ is shown in Fig.~\ref{fig:I_L_stat}(e). Similarly, the the average right current $I_R$ and its fluctuations $S_R$ are plotted in Fig.~\ref{fig:I_L_stat}(b) and Fig.~\ref{fig:I_L_stat}(f), respectively.
From the colormaps of $I_L$ and $I_R$, we observe that their dependence on $\kappa$ and $\Gamma$ differs when $\Gamma\neq \kappa$. 
For the strong $\Gamma$ or $\kappa$ values, the current and noise decrease which is a strong coupling effect and a reminder of the quantum Zeno effect.
Similar to $I_L$ and $I_R$, the values of $S_L$ and $S_R$ are also different for arbitrary $\Gamma$ and $\kappa$ values. The positive values of $I_L$ ($I_R$) indicate that the net current is coming into the system from the left (right) end and the negative values indicate that the net current is going out of the system from the left (right) end. It is important to note that such negative sign in the currents does not imply the emergence of nonreciprocity in the lattice. This negativity is solely due to unequal gain and loss strength which results unequal transmission functions of fermions from both the end sites to the gain-loss channels. However, along the line $\Gamma=\kappa$ (black dashed line in Fig.~\ref{fig:I_L_stat}) where the three conditions~\ref{cond:A}-\ref{cond:C} of identical statistics of $\mathcal{I}_L$ and $\mathcal{I}_R$ are fulfilled, $I_L$ and $I_R$ are equal in this case with opposite sign . To appreciate this equality, in Fig.~\ref{fig:I_L_stat}(c) we have plotted the total bulk current coming out of the system, $I_{\rm bulk}=I_L+I_R$ which is zero when $I_L=-I_R$ i.e., along the $\Gamma=\kappa$ line. Above this line, the loss of fermions is more than gain and hence, $I_{\rm bulk}>0$ and below this line, the rate for gaining fermions is more than the loss and hence $I_{\rm bulk}<0$. The current fluctuations $S_L$ and $S_R$ are also equal along this line. 
In Fig.~\ref{fig:I_L_stat}(d), we have plotted
the net left to right current $I_{LR}$ defined in Eq.~\eqref{eq:left_to_right_current} which is always positive for our choice of parameters
and hence confirms the net current flows from left to the right direction. 

In the next subsection, we analyze the correlated gain-loss case, where we will discuss the emergence of non-reciprocity in the setup and its impact in current statistics.
\begin{figure*}
    \centering
    \includegraphics[width=0.5\columnwidth]{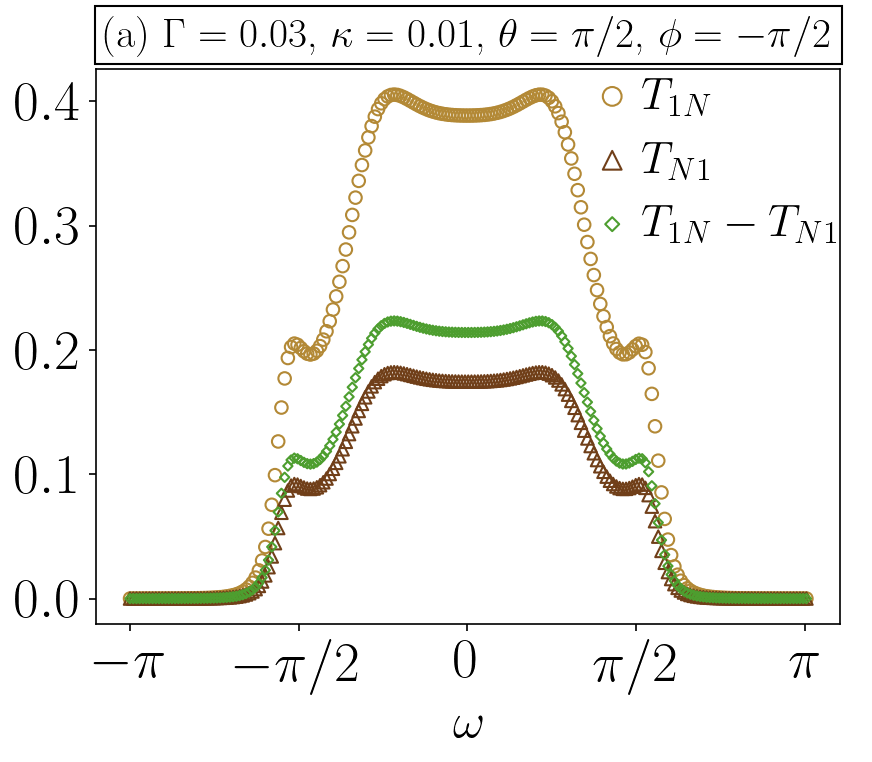}%
    \includegraphics[width=0.5\columnwidth]{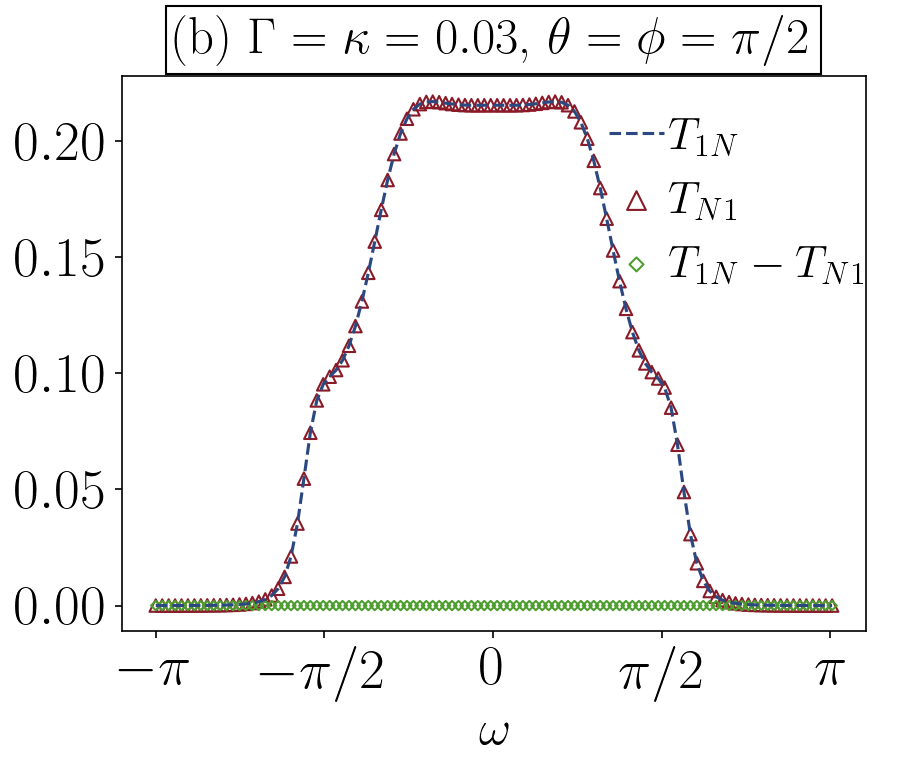}%
    \includegraphics[width=0.5\columnwidth]{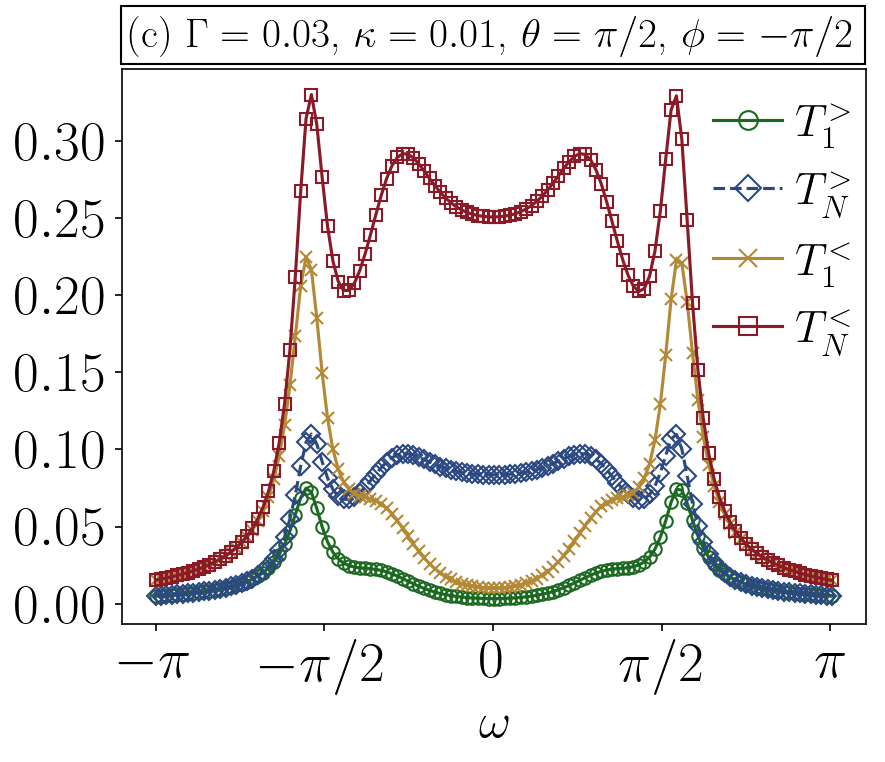}%
    \includegraphics[width=0.5\columnwidth]{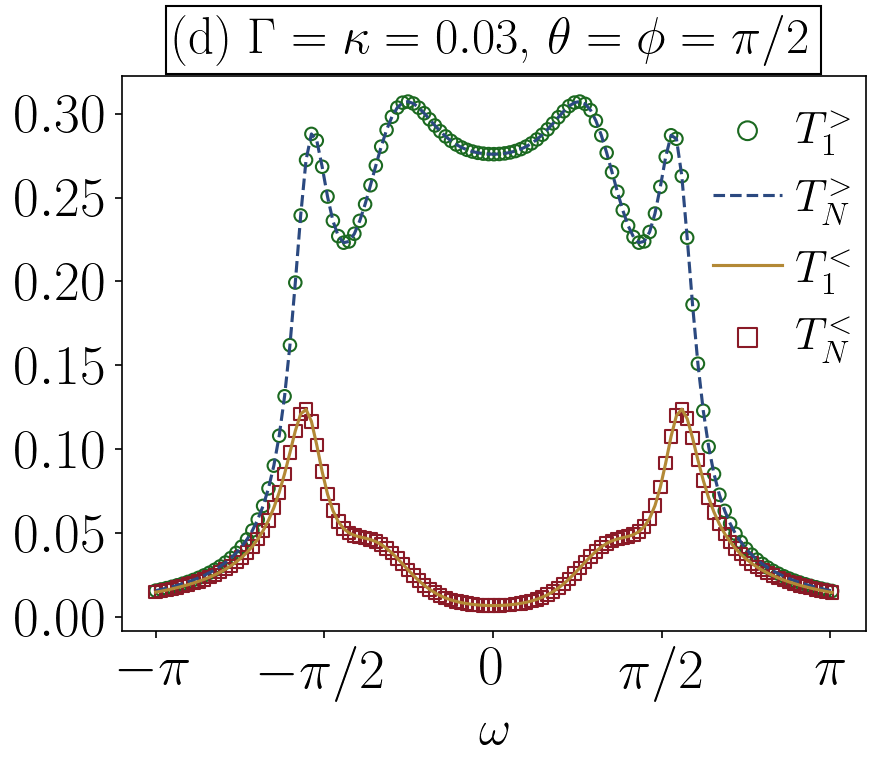}
    \caption{Correlated gain-loss case: plot of different transmission functions in presence of correlated gain-loss channels. We set $\alpha_1=\beta_N=1.0$, $\alpha_N=\beta_1=0.5$ which satisfy the condition (i) $\alpha_1+\beta_1=\alpha_N+\beta_N$ and (iv) $\alpha_1=\beta_N$ and $\alpha_N=\beta_1$. The system size is chosen to be $N=6$. (a) Left to right transmission function $T_{1N}(\omega)$ [Eq.~\eqref{eq:transmission_gainloss}] and right to left transmission function $T_{N1}(\omega)$ [Eq.~\eqref{eq:transmission_gainloss}] are plotted as a function of frequency $\omega$ when $\Gamma\neq \kappa$ and $\theta\neq \phi$ and they are unequal. This inequality in the transmission functions is referred as nonreciprocity in the system which can also be guaranteed by the plot of $T_{1N}(\omega)-T_{N1}(\omega)$ which is evidently nonzero. (b) $T_{1N}(\omega)$ and $T_{N1}(\omega)$ become equal when we set $\Gamma=\kappa$ and $\theta=\phi$ which thereby ensures reciprocity in the system. (c) The plot of the transmission functions from the $1$st to the gain and loss channels i.e., $T_{1}^{<}(\omega)$ and $T_{1}^{>}(\omega)$ and from the $N$-th site to the gain and loss channels i.e., $T_N^{<}(\omega)$ and $T_{N}^{>}(\omega)$ are plotted for the same set of parameters as of (a) when the system is nonreciprocal and hence we can see that they are all unequal. One we impose $\Gamma=\kappa$ and $\theta=\phi$, the reciprocity ensures $T_{1}^{>}(\omega)=T_{N}^{<}(\omega)$ and $T^{<}_1(\omega)=T_{N}^{>}(\omega)$.}
    \label{fig:corr_transmission}
\end{figure*}
\subsection{Correlated gain-loss channels}
\label{subsec:correlated_gainloss}
\begin{figure*}
    \centering
    \includegraphics[width=6.0cm]{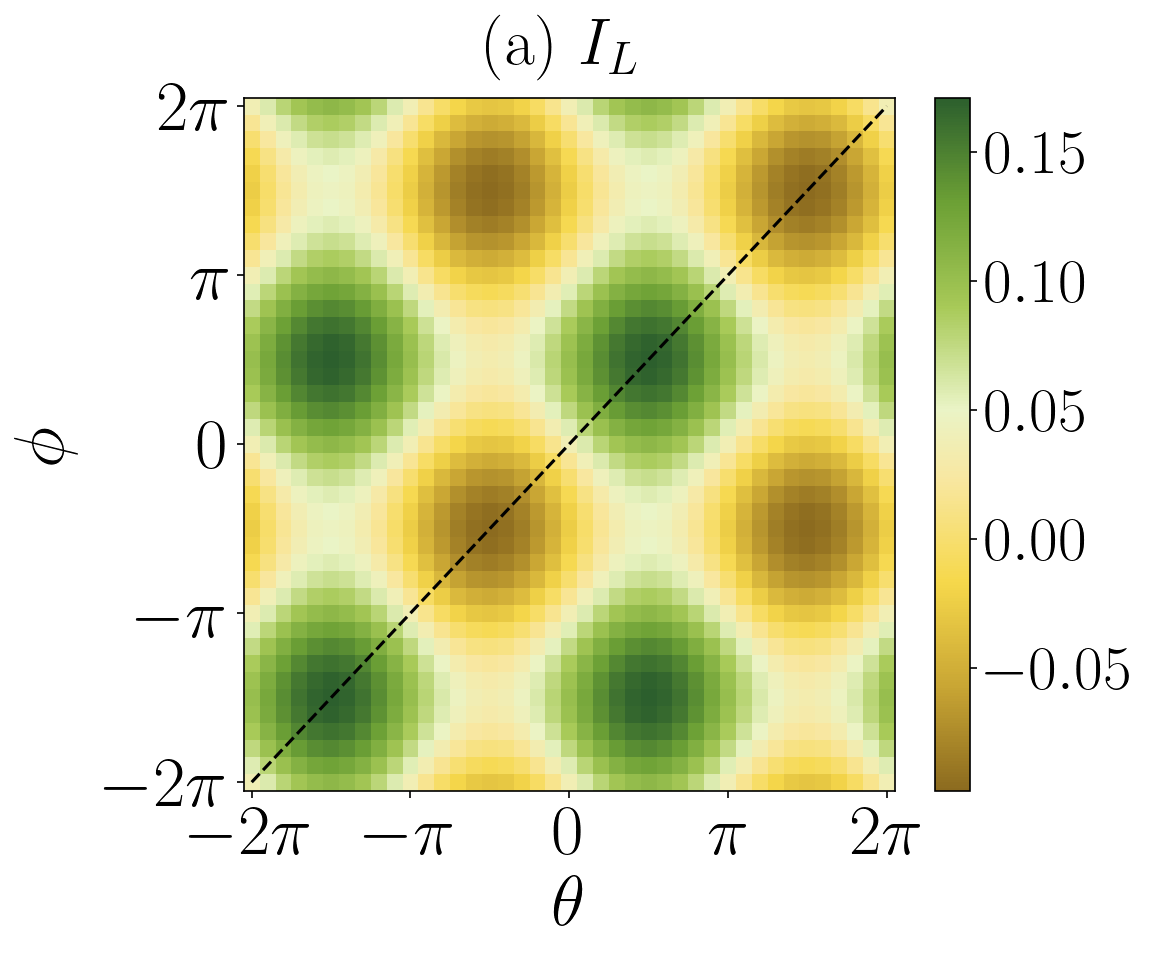}%
    \includegraphics[width=6.0cm]{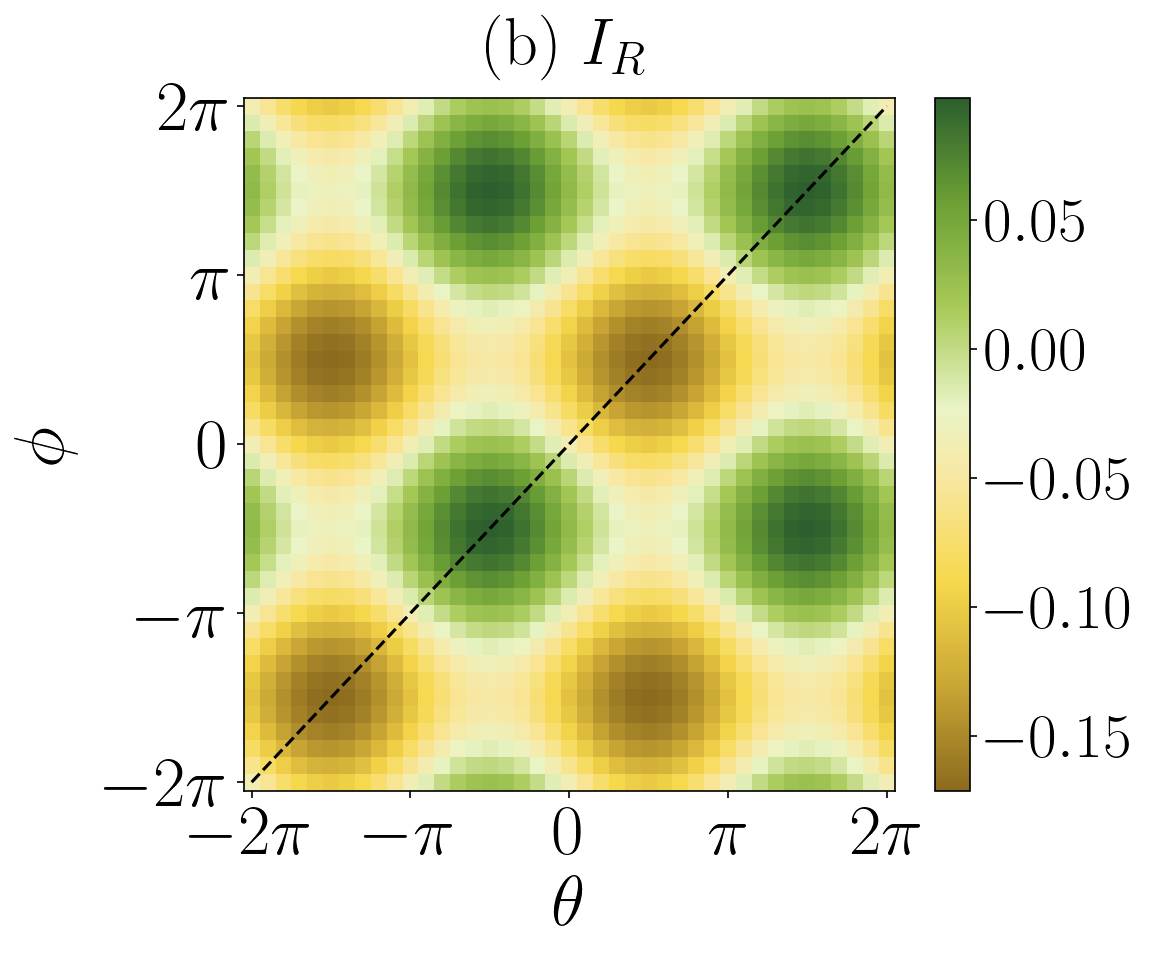}%
    \includegraphics[width=6.0cm]{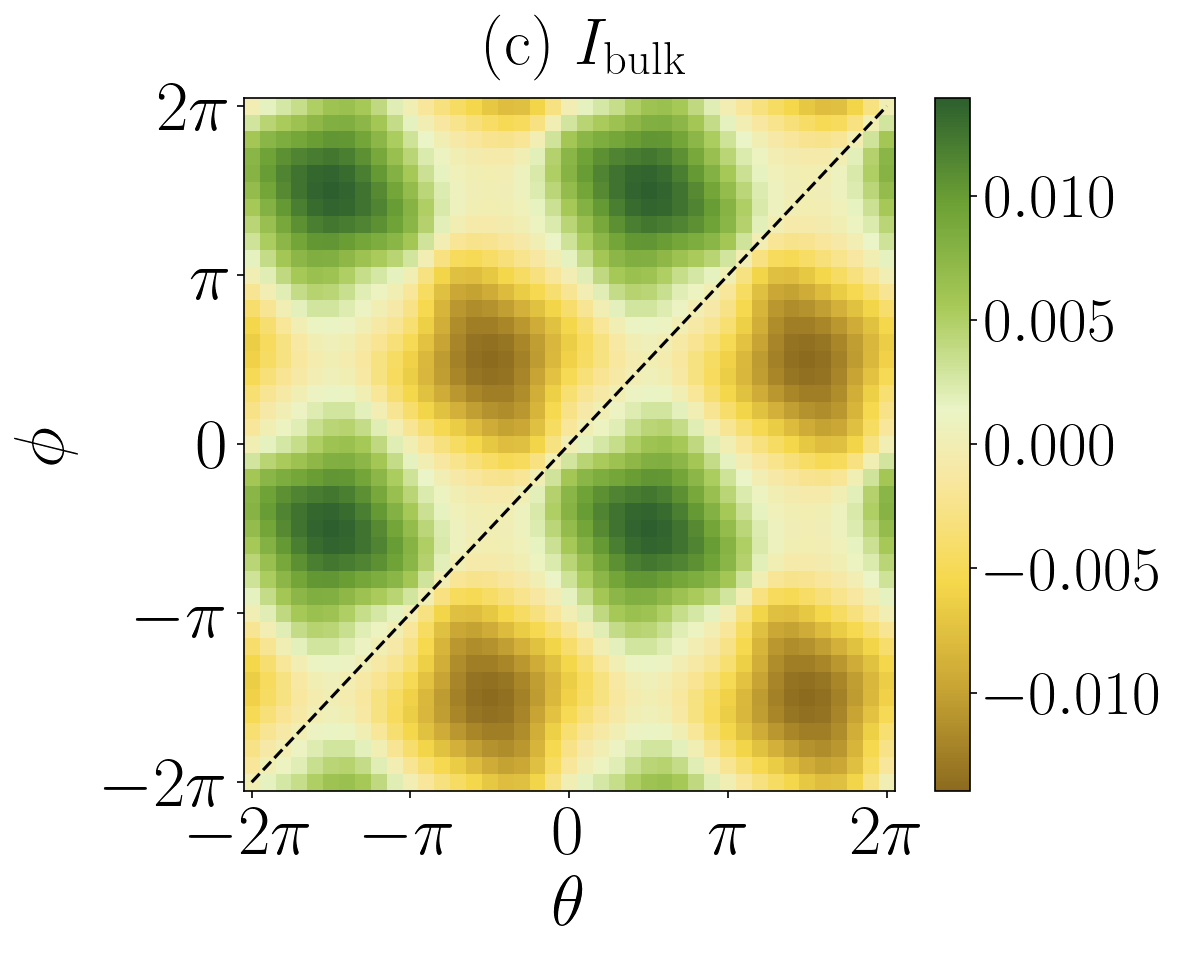}
    \includegraphics[width=6.0cm]{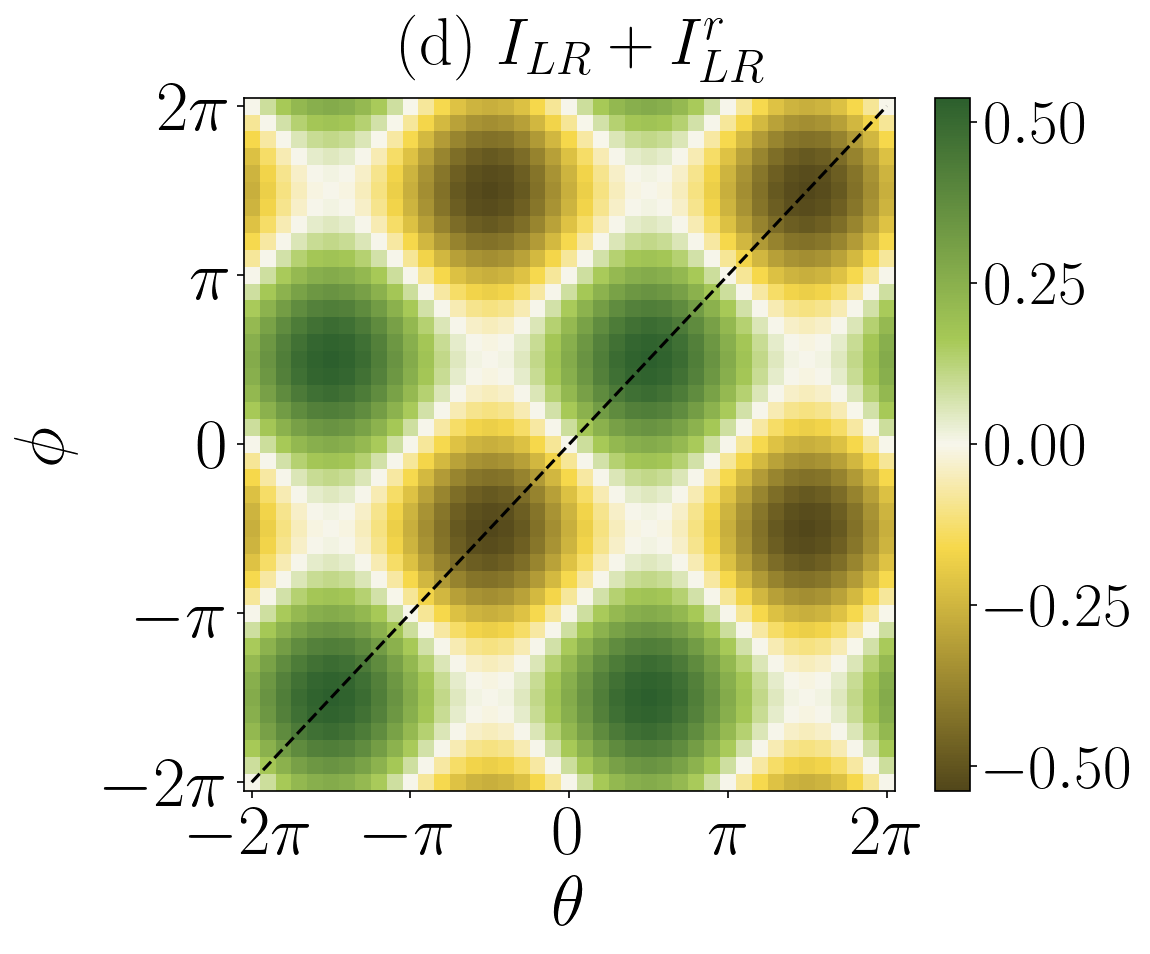}%
    \includegraphics[width=6.0cm]{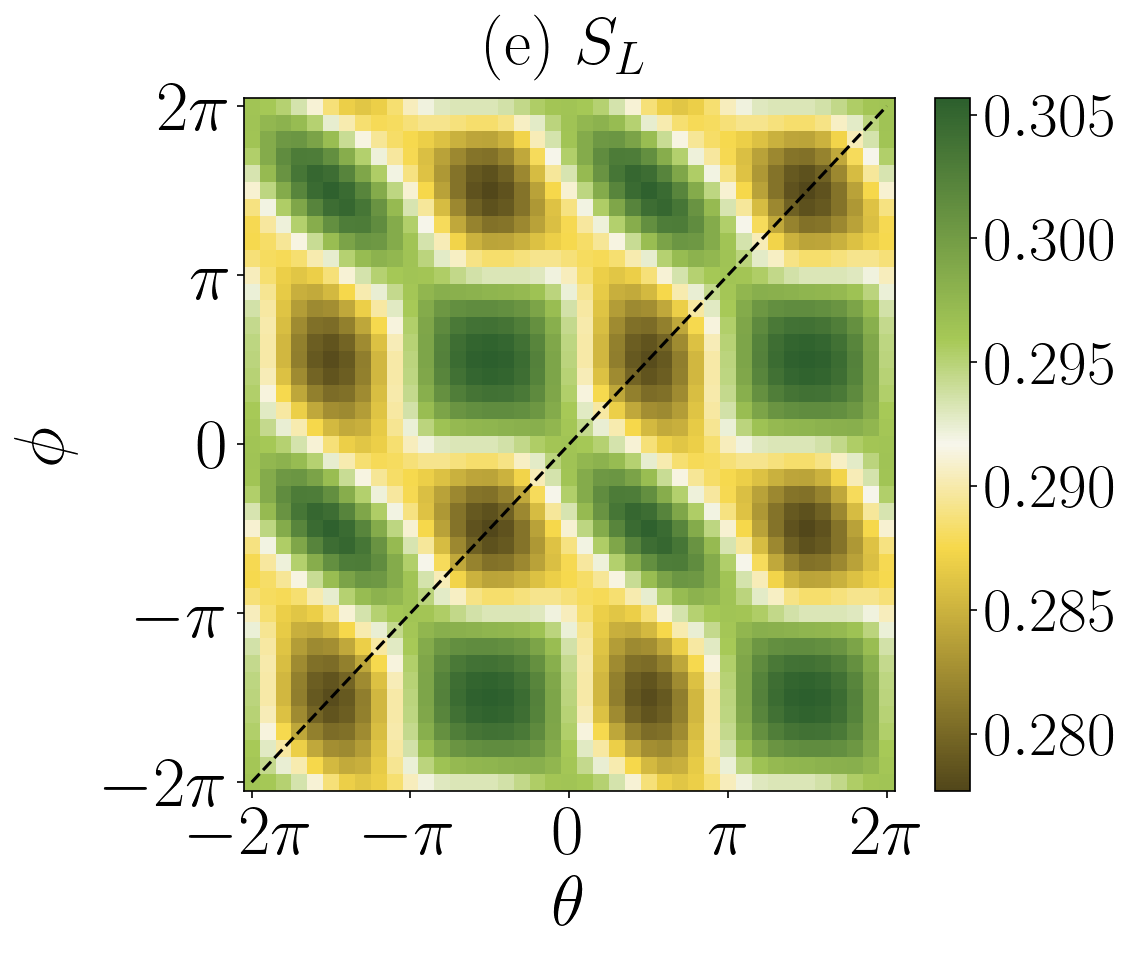}%
    \includegraphics[width=6.0cm]{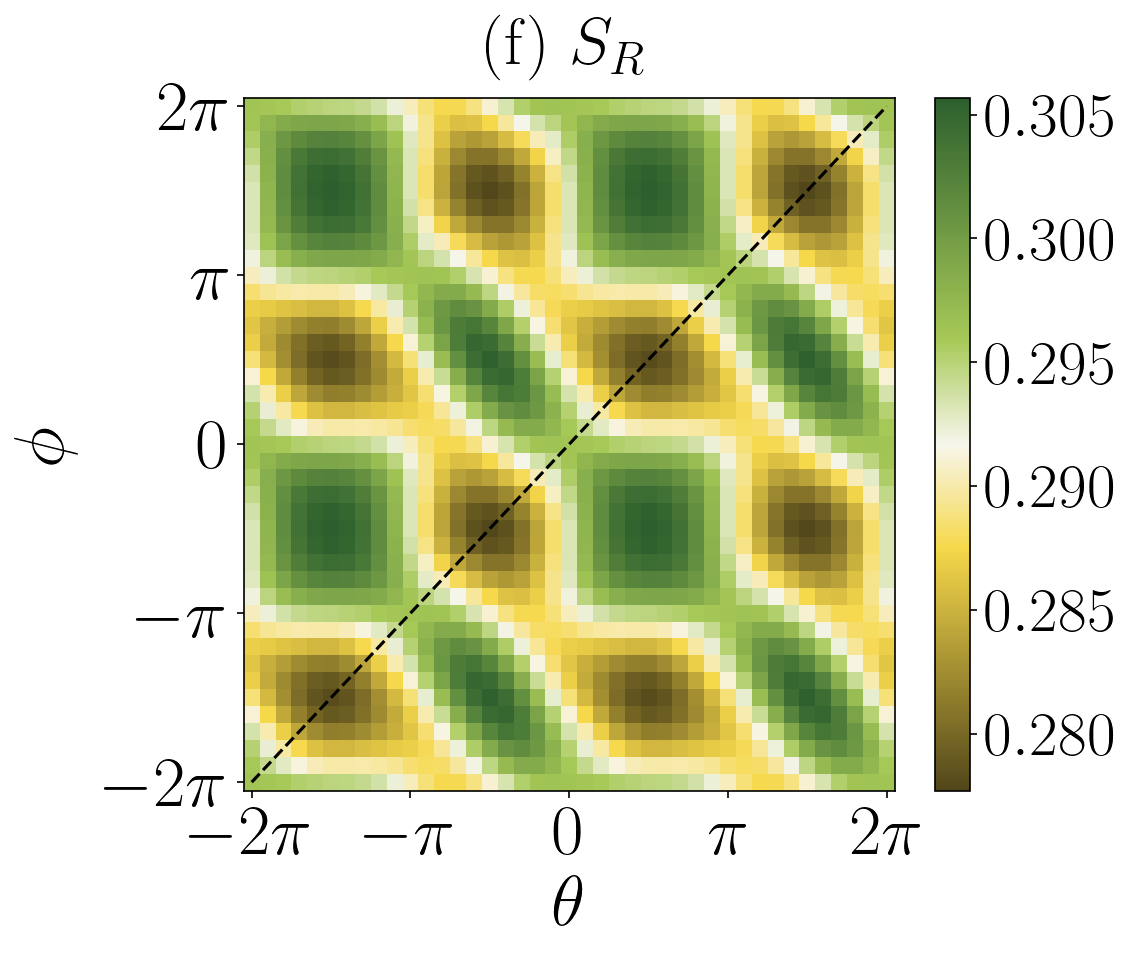}
    \caption{Correlated gain-loss case: colormap plot of the statistics of left and right current in correlated gain-loss case. The parameters chosen here are $\alpha_1=\beta_N=1.1$, $\alpha_N=\beta_1=0.9$, $\Gamma=\kappa=0.05$. (a) Plot of average left current $I_L$ using Eq.~\eqref{eq:I_L_expr}, (b) average right current $I_R$ using Eq.~\eqref{eq:I_R_expr}. The values of $I_L$ and $I_R$ are in general different for general $\theta$, $\phi$ values although both are periodic with $\theta$ and $\phi$. Along the line $\theta=\phi$ (black dashed line) where the reciprocity in the system is revived, $I_L$ and $I_R$ are equal in magnitude and opposite in sign. (c) Plot of the total bulk current going out of the system, $I_{\rm bulk}$ defined in Eq.~\eqref{eq:bulk_current} which is zero along $\theta=\phi$ because $I_L=-I_R$. (d) Plot of the current rectification factor $I_{LR}+I_{LR}^{r}$ where $I_{LR}$ is the net left to right current in the system under the boundary rates $\alpha_1$, $\beta_1$, $\alpha_N$, and $\beta_N$ and $I_{LR}^{r}$ is the net left to right current under interchanging the bias. As $I_{LR}+I_{LR}^{r}$ is nonzero, it implies diode effect in the setup. (e)-(f) Plot of noise $S_L$ and $S_R$ defined in Eq.~\eqref{eq:S_L} and~\eqref{eq:S_R} corresponding to the current $\mathcal{I}_L$ and $\mathcal{I}_R$ respectively. Although they are periodic with $\theta$ and $\phi$, but their dependence on $\theta$, $\phi$ are evidently different. Along the $\theta=\phi$ line where the CGF of left and right current coincide, $S_L$ and $S_R$ are also same. }
    \label{fig:S_vs_theta}
\end{figure*}
In this subsection, we discuss the case of correlated gain-loss channels in the bulk. In particular, we consider the choice  $\Gamma_1=\Gamma_2=\Gamma$ and $\kappa_1=\kappa_2=\kappa$ in Eq.~\eqref{eq:AB_def} without any loss of generality. In this case, the bulk self energies $\gamma_{\rm bulk}^{>,<,R,K}$ are asymmetric tridiagonal matrices. As a result, $|G^{R}_{1N}(\omega)|\neq |G^{R}_{N1}(\omega)|$ which can be shown following the transfer matrix approach~\cite{loss2024}. Accordingly, the transmission functions $T_{1N}(\omega)$ and $T_{N1}(\omega)$ are in general unequal for arbitrary values of $\Gamma$, $\kappa$, $\theta$, and $\phi$. 
This behaviour is in sharp contrast to the local gain-loss setup discussed in sec.~\ref{subsec:localloss} where $T_{1N}(\omega)$ and $T_{N1}(\omega)$ are always equal irrespective of the choice of parameters. The unequal transmission functions in the correlated gain-loss case render the system {\it nonreciprocal}~\cite{Clerk2015,PhysRevLett.130.110401,schiro2025}, as illustrated in  Fig.~\ref{fig:corr_transmission}(a). A direct consequence of this nonreciprocity is that the other transmission functions $T_{1,N}^{>,<}(\omega)$ are also generally unequal and remain so even when all conditions~\ref{cond:A}-\ref{cond:C} from the local gain-loss case are imposed. This is shown in Fig.~\ref{fig:corr_transmission}(c). However, the reciprocity can be restored 
in presence of correlated gain-loss channels by additionally imposing the condition $\theta=\phi$, along with the three conditions derived for the local gain-loss case. The combined constraints  $\theta=\phi$ along with $\Gamma=\kappa$ results in symmetric tridiagonal self energies which lead to  $T_{1N}(\omega)=T_{N1}(\omega)$ [see Fig.~\ref{fig:corr_transmission}(b)] and $T_{1}^{>}(\omega)=T_{N}^{<}(\omega)$, $T^{>}_N(\omega)=T_{N}^{<}(\omega)$. Thus, for the correlated gain-loss case configuration, the following four conditions are required to obtain identical statistics of $\mathcal{I}_L$ and $\mathcal{I}_R$,
\begin{enumerate}[label=\roman*., ref=(\roman*)]
    \item $\alpha_1+\beta_1=\alpha_N+\beta_N$, \label{cond_1}
    \item $\Gamma=\kappa$, \label{cond_2}
    \item $\theta=\phi$, \label{cond_3}
    \item $\alpha_1=\beta_N$ and $\alpha_N=\beta_1$. \label{cond_4}
\end{enumerate}
Here, the third condition is an additional constraint compared to the local gain-loss case. Once these four conditions are enforced, a balanced gain-loss scenario is once again emerges in the system, and as a result, the system respects $\mathcal{PT}$ symmetry. Given these conditions, if we now interchange the boundary rates ($\alpha_1\leftrightarrow\alpha_N$ and $\beta_1\leftrightarrow\beta_N$), the magnitude of the current does not become identical even though $T_{1N}(\omega)=T_{N1}(\omega)$. This is because unlike the local gain-loss case, the other four transmission functions i.e., $T_{1,(N)}^{>,<}(\omega)$ are not all the same. 
In this sense the correlated gain-loss setup is {\it nonreciprocal} and hence diode effect can be observed.

We now illustrate in Fig.~\ref{fig:S_vs_theta} the current statistics for the correlated gain-loss case by plotting the first two cumulants i.e., the average current $I_L$ and $I_R$ [Eq.~\eqref{eq:I_L_expr}
and~\eqref{eq:I_R_expr}, respectively],  and current fluctuations $S_L$ and $S_R$ [Eq.~\eqref{eq:S_L}
and~\eqref{eq:S_R}, respectively]. In our numerical analysis, we enforce only conditions~\ref{cond_1}, \ref{cond_2}, and~\ref{cond_4} by choosing the parameters $\alpha_1=\beta_N=1.1$, $\alpha_N=\beta_1=0.9$, and $\Gamma=\kappa=0.05$, while varying the phases $\theta$ and $\phi$.
From Fig.~\ref{fig:S_vs_theta}(a) and (b), we observe the periodic dependence of $I_L$ and $I_R$ on the phases $\theta$ and $\phi$. For general $\theta$, $\phi$ values, $I_L$ and $I_R$ are different. However, along the line $\theta=\phi$, where all the four conditions are fulfilled, the reciprocity is revived and the currents are equal. The equality can also be inferred from the Fig.~\ref{fig:S_vs_theta}(c), where the bulk current $I_{\rm bulk}=I_L+I_R$ is plotted and it vanishes along the line $\theta=\phi$. Another interesting scenario in the correlated gain-loss case is the emergence of diode effect in the setup. In Fig.~\ref{fig:S_vs_theta}(d), we have plotted the current rectification factor $I_{LR}+I_{LR}^{r}$. Here $I_{LR}$ defined in Eq.~\eqref{eq:left_to_right_current} is the net left to right current flowing through the system under the boundary drives by the rates $\alpha_1$, $\beta_1$, $\alpha_N$, and $\beta_N$ whereas $I_{LR}^{r}$ is the net left to right current under interchanging the boundary rates $\alpha_1\leftrightarrow\alpha_N$ and $\beta_1\leftrightarrow\beta_N$. The nonzero values of the rectification factor implies that the setup exhibits diode effect. Such current rectification or diode like effect is an interesting phenomenon in correlated gain-loss setup. We further plot the current fluctuations $S_L$ and $S_R$ which also have very distinct dependence on $\theta$ and $\phi$ in general. However, for $\theta=\phi$, the noise at both the boundaries are again same as the entire CGF of $\mathcal{I}_L$ and $\mathcal{I}_R$ are identical along this line.

\section{Summary and outlook} \label{summary}
In this work, we have studied the full counting statistics of particle currents in a boundary driven free fermionic lattice subjected to correlated gain-loss dissipators across the entire lattice. 
The dynamics of the system was modeled using GKSL master equation. Employing the Feynman path integral approach, we have analytically derive the CGF. 

We first analyzed the case with only boundary drives, in absence of any gain-loss channels and the resulting CGF takes the form which is analogous to the well-known Levitov-Lesovik formula. We then incorporated gain-loss channels into the setup and obtained the corresponding CGF for the left and right particle currents. In the presence of such gain-loss channels, the current statistics at the two boundaries are, in general, different except in a balanced gain-loss scenario when $\mathcal{PT}$ symmetry is restored and statistics become identical. Furthermore, when correlated gain-loss is introduced, the system develops nonreciprocity in the transmission function, leading to a diode like behaviour in the presence of boundary drives. Such a behaviour is completely absent in presence of local gain-loss channels and purely a consequence of correlated dissipator. Our study therefore unravel the importance of engineered dissipators to obtain interesting functionalities that are important to develop quantum technologies.

Future directions could involve 
extending the path integral framework to study driven-dissipative systems in presence of Lindblad dissipators. Our approach can  be adapted for boundary driven systems subjected to dephasing mechanisms in the bulk of the lattice.

\section*{Acknowledgements}  
 KG acknowledges Abhishek Dhar, Sandipan Manna for useful discussions. BKA acknowledges the CRG grant No. CRG/2023/003377 from ANRF, Government of India. KG would like to acknowledge the Prime Minister's Research Fellowship (ID- 0703043), Government of India for funding. KG and BKA acknowledge the National Supercomputing Mission (NSM) for providing computing resources of ‘PARAM Brahma’ at IISER Pune, which is implemented by C-DAC and supported by the Ministry of Electronics and Information Technology (MeitY) and DST, Government of India.

\onecolumngrid
\appendix

\setcounter{figure}{0}
\renewcommand{\thefigure}{A\arabic{figure}}

\begin{center}
{\textbf{\underline{Appendix}}}
\end{center}

\section{Derivation of the Keldysh action in Eq.~\eqref{action} for case-1: only boundary drives}
\label{app:action}
\begin{figure}[h!]
    \centering
    \includegraphics[width=\linewidth]{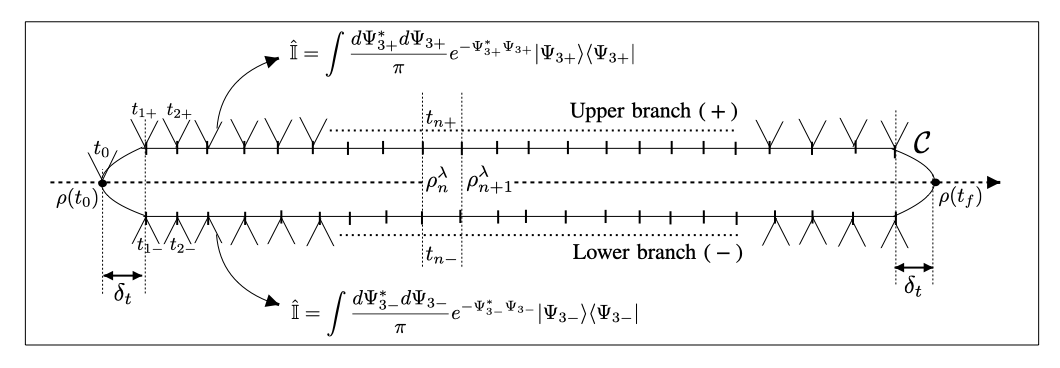}
    \caption{Schematic of the Keldysh contour describes the evolution of a mixed state $\rho$ from initial time $t_0$ to a final time $t_f$. This picturizes how the evolution is described in coherent basis by the insertion of identity. The Keldysh contour has two branches: upper ($+$) branch corresponding to the forward evolution and lower ($-$) branch corresponding to the backward evolution. In each branch the time is discretized in small intervals $\delta_t$ and in each intervals the resolution of identity $\hat{\mathbb{I}}$ in terms of coherent basis is inserted. }
    \label{fig:Keldysh_contour}
\end{figure}

In this appendix, we derive the moment generating function (MGF) corresponding to the statistics of integrated particle current $\mathcal{N}_L$ from left the end of the lattice. The definition of the moment generating function is given as,
\begin{align}
    \mathcal{Z}(\lambda)=\rm Tr\Big[e^{(t_f - t_0)\mathcal{L}_\lambda}\rho(t_0)\Big], \label{def_MGF_supp}
\end{align}
where $\lambda$ is the counting field and $\mathcal{L}_{\lambda}$ is the Liouvillian superoperator with the jump term dressed by the counting field $\lambda$ defined in the main text~\eqref{Lindblad_eq_counting_field}.
We use the Schwinger-Keldysh path integral formalism and express the MGF in Eq.~\eqref{def_MGF_supp} in terms of path integral by going to the coherent basis picture for the fermions. The coherent basis for the fermions is defined using the Grasmann variables as the following~\cite{kamenev2011field,Sieberer_2016},
\begin{align}
    c_{i}|\psi_i\rangle=\psi_{i}|\psi_i\ra ,\quad i = 1,2,\dots,N
\end{align}
where $\psi_i$'s correspond to the eigenvalues of the anhiliation operator $c_{i}$'s with the corresponding eigenstates forming coherent basis $|\psi_i\rangle$. Here $N$ is the number of lattice sites. $\psi_i$ satisfies the anticommutation relation,
\begin{align}
    \psi_i \psi_j+\psi_j\psi_i=0,\quad \psi_i^2=0,\quad \psi_i^{*}\psi_j+\psi_j^{*}\psi_i=\delta_{ij}.
\end{align}
The coherent basis is not orthogonal $\big(\la \psi_i|\phi_j\rangle=e^{\psi^{*}_i\phi_j}\big)$ and also it is overcomplete. The completeness relation is,
\begin{align}
    \int \frac{d\psi^{*}d\psi}{\pi}e^{-|\psi|^{2}}|\psi\rangle\la\psi|=\mathbb{I}. \label{eq:identity_coherent_supp}
\end{align}
We now express the Eq.~\eqref{def_MGF_supp} in the coherent basis. As the evolution of the density matrix $\rho$ by the Liouvellian superoperator $\mathcal{L}_\lambda$ is best described using Keldysh contour, we have presented a schematic of the Keldysh contour in Fig.~\ref{fig:Keldysh_contour}. The Fig.~\ref{fig:Keldysh_contour} is also useful for the visual representation of the procedure to obtain the MGF in the form of a Path integral. In this procedure, we first discretize the time by dividing the full evolution time i.e., $t_f - t_0$ into small time intervals of length $\delta_t$ and write the $\lambda$-dressed density matrix at $n$-th time instant as,
\begin{align}
\rho^{\lambda}_{n+1}=\big[e^{\delta_t\mathcal{L}_{\lambda}}\big]^{n+1}\rho(t_0)=\big[\mathbb{I}+\delta_t\mathcal{L}_{\lambda}\big]^{n+1}\rho(t_0)= \big[\mathbb{I}+\delta_t\mathcal{L}_{\lambda}\big]\rho^{\lambda}_{n}.\label{rho_n_supp}
\end{align}
For the path integral representation of the MGF $\mathcal{Z}(\lambda)$, it is required to represent $\rho_n$ in the coherent basis and hence in the Eq.~\eqref{rho_n_supp}, we further insert the resolution of identity for the coherent states [Eq.~\eqref{eq:identity_coherent_supp}] at each instant of time $t_n$. The procedure of inserting identities along with the time evolution of the density matrix is pictorially represented in Fig.~\ref{fig:Keldysh_contour}. The Keldysh contour has an upper ($+$) and lower ($-$) branch which represent the forward (operators acting from the left side of $\rho$) and backward (operators acting from the right side of $\rho$) time evolution of the density matrix, respectively. Once the time is discretized, the full time evolution is represented by $(t_f-t_0)/\delta_t$ number of steps, and after each step of evolution of $\rho$, the resolution of identities are introduced from the left and right side of $\rho$ with corresponding branch indices [see Fig.~\ref{fig:Keldysh_contour}].
Hence, for the forward evolution, we insert the identity involving coherent states $|\Psi_{n+}\ra$ from the left side and similarly for the backward evolution, identity involving $|\Psi_{n-}\ra$ is inserted from the right side of the density matrix $\rho^{\lambda}_n$. The dressed density matrix $\rho^{\lambda}_n$ in coherent basis then takes the form,
\begin{align}
    \rho^{\lambda}_n=\int \frac{d\Psi^{*}_{n+}d\Psi_{n+}}{\pi}\frac{d\Psi^{*}_{n-}d\Psi_{n-}}{\pi}e^{-(\Psi^{*}_{n+}\Psi_{n+}+\Psi^{*}_{n-}\Psi_{n-})}|\Psi_{n+}\rangle\la\Psi_{n+}|\rho^{\lambda}_n|\Psi_{n-}\ra\la\Psi_{n-}|,\label{coherent_basis_rep_supp}
\end{align}
where $|\Psi_{n,\pm}\ra=|\psi^{1}_{n,\pm}\rangle|\psi^{2}_{n,\pm}\rangle\dots|\psi^{N}_{n,\pm}\rangle$ with the superscript representing the lattice site index. Combining Eq.~\eqref{rho_n_supp} and Eq.~\eqref{coherent_basis_rep_supp}, we obtain the state at time $\rho_{n+1}$,
\begin{align}
    \rho^{\lambda}_{n+1}=\int& \frac{d\Psi^{*}_{n+}d\Psi_{n+}}{\pi}\frac{d\Psi^{*}_{n-}d\Psi_{n-}}{\pi}e^{-(\Psi^{*}_{n+}\Psi_{n+}+\Psi^{*}_{n-}\Psi_{n-})}|\Psi_{n+}\rangle\la\Psi_{n-}|\la\Psi_{n+}|\rho^{\lambda}_n|\Psi_{n-}\ra\nonumber\\&+\delta_t\int \frac{d\Psi^{*}_{n+}d\Psi_{n+}}{\pi}\frac{d\Psi^{*}_{n-}d\Psi_{n-}}{\pi}e^{-(\Psi^{*}_{n+}\Psi_{n+}+\Psi^{*}_{n-}\Psi_{n-})}\mathcal{L}_{\lambda}\Big[|\Psi_{n+}\rangle\la\Psi_{n-}|\Big]\la\Psi_{n+}|\rho^{\lambda}_n|\Psi_{n-}\ra+O(\delta_t^2), \label{r_n1_supp}
\end{align}
where recall that the expression of the dressed Liouvillian $\mathcal{L}_{\lambda}$ is,
\begin{align}
    \mathcal{L}_\lambda\,\bullet=-iH\bullet + \bullet iH+\mathcal{D}^{\lambda}_1[\bullet]+\mathcal{D}_N[\bullet].\label{liou_boundry}
\end{align}
Here $\mathcal{D}^{\lambda}_1[\bullet]=\alpha_{1}\big[2e^{i\lambda} c^{\dagger}_{1}\bullet c_{1}-\{c_{1}c^{\dagger}_{1},\bullet\} \big]+\beta_{1}\big[2e^{-i\lambda}c_{1}\bullet c^{\dagger}_{1}-\{c^{\dagger}_{1}c_{1},\bullet\} \big]$ and $\mathcal{D}_{N}=\alpha_{N}\big[2 c^{\dagger}_{N}\bullet c_{N}-\{c_{N}c^{\dagger}_{N},\bullet\} \big]+\beta_{N}\big[2c_{N}\bullet c^{\dagger}_{N}-\{c^{\dagger}_{N}c_{N},\bullet\} \big]$. Next, we need to represent $\rho_{n+1}^{\lambda}$ in coherent basis and hence we need to obtain $\la \Psi_{(n+1)+}|\rho_{n+1}^{\lambda}|\Psi_{(n+1)-}\ra$.
To obtain this, we need to calculate $\la \Psi_{(n+1)+}|\mathcal{L}_{\lambda}\big[|\Psi_{n+}\ra\la\Psi_{n-}|\big]|\Psi_{(n+1)-}\ra$, which we write as the following,
\begin{align}
\la\Psi_{(n+1)+}|\mathcal{L}_\lambda\,\big[|\Psi_{n+}\ra\la\Psi_{n-}|\big]|\Psi_{(n+1)-}\ra&=f_\lambda(\Psi^{*}_{(n+1)+},\Psi_{n+},\Psi^{*}_{n-},\Psi_{(n+1)-})\la\Psi_{(n+1)+}|\Psi_{n+}\ra\la\Psi_{n-}|\Psi_{(n+1)-}\ra.\label{liouvel_coherent_supp}
\end{align}
We can break $f_\lambda(\Psi^{*}_{(n+1)+},\Psi_{n+},\Psi^{*}_{n-},\Psi_{(n+1)-})$ in three terms. The first term is the contribution from the Hamiltonian, $f_H(\Psi^{*}_{(n+1)+},\Psi_{n+},\Psi^{*}_{n-},\Psi_{(n+1)-})$ which can be obtained as,
\begin{align}
    -i\Big(\la\Psi_{(n+1)+}| &H|\Psi_{n+}\ra\la\Psi_{n-}|\Psi_{(n+1)-}\ra - \la\Psi_{(n+1)+}|\Psi_{n+}\ra\la\Psi_{n-}| H|\Psi_{(n+1)-}\ra\Big)\nonumber\\&=-i\sum_{i,j=1}^{N}h_{ij}[\psi_{(n+1)+}^{i*}\psi^{j}_{n+}-\psi_{n-}^{i*}\psi^{j}_{(n+1)-}] \,\la\Psi_{(n+1)+}|\Psi_{n+}\ra\la\Psi_{n-}|\Psi_{(n+1)-}\ra,\label{hamiltonian_coh_supp}
\end{align}
where we identify,
\begin{align}
    f_H(\Psi^{*}_{(n+1)+},\Psi_{n+},\Psi^{*}_{n-},\Psi_{(n+1)-})=-i\sum_{i,j=1}^{N}h_{ij}[\psi_{(n+1)+}^{i*}\psi^{j}_{n+}-\psi_{n-}^{i*}\psi^{j}_{(n+1)-}] \label{fH_supp_boundary}
\end{align}
Similarly, we obtain the dissipative terms $f^{\lambda}_1(\Psi^{*}_{(n+1)+},\Psi_{n+},\Psi^{*}_{n-},\Psi_{(n+1)-})$ and $f_N(\Psi^{*}_{(n+1)+},\Psi_{n+},\Psi^{*}_{(n+1)-},\Psi_{n-})$ in the Eq.~\eqref{liouvel_coherent_supp} by evaluating $\la\Psi_{(n+1)+}|\mathcal{D}^{\lambda}_1\big[|\Psi_{n+}\ra\la\Psi_{n-}|\big]|\Psi_{(n+1)-}\ra$ and $\la\Psi_{(n+1)+}|\mathcal{D}_N\big[|\Psi_{n+}\ra\la\Psi_{n-}|\big]|\Psi_{(n+1)-}\ra$ respectively and this we obtain,
\begin{align}
    f^{\lambda}_1(\Psi^{*}_{(n+1)+},\Psi_{n+},\Psi^{*}_{n-},\Psi_{(n+1)-})&=\alpha_1[2e^{i\lambda}\psi^{1*}_{(n+1)+}\psi_{(n+1)-}^{1}+\psi^{1*}_{(n+1)+}\psi_{n+}^{1}+\psi^{1*}_{n-}\psi_{(n+1)-}^{1}]\nonumber\\&+\beta_1[2e^{-i\lambda}\psi^{1}_{n+}\psi_{n-}^{1*}-\psi^{1*}_{(n+1)+}\psi_{n+}^{1}-\psi^{1*}_{n-}\psi_{(n+1)-}^{1}]\label{f1_supp}\\
    f_N(\Psi^{*}_{(n+1)+},\Psi_{n+},\Psi^{*}_{n-},\Psi_{(n+1)-})&=\alpha_N[2\psi^{N*}_{(n+1)+}\psi_{(n+1)-}^{N}+\psi^{N*}_{(n+1)+}\psi_{n+}^{N}+\psi^{N*}_{n-}\psi_{(n+1)-}^{N}]\nonumber\\&+\beta_N[2\psi^{N}_{n+}\psi_{n-}^{N*}-\psi^{N*}_{(n+1)+}\psi_{n+}^{N}-\psi^{N*}_{n-}\psi_{(n+1)-}^{N}].\label{fN_supp}
\end{align}
Combining Eq.~\eqref{fH_supp_boundary},~\eqref{f1_supp}, and~\eqref{fN_supp}, one can obtain the full expression of $\la\Psi_{(n+1)+}|\mathcal{L}_\lambda\big[|\Psi_{n+}\ra\la\Psi_{n+}|\big]|\Psi_{(n+1)-}\ra$. Using that, finally we write $\la\Psi_{(n+1)+}|\rho^{\lambda}_{n+1}|\Psi_{(n+1)-}\ra$ as, 
\begin{align}
 \la\Psi_{(n+1)+}|\rho^{\lambda}_{n+1}|\Psi_{(n+1)-}\ra=\int \frac{d\Psi^{*}_{n+}d\Psi_{n+}}{\pi}\frac{d\Psi^{*}_{n-}d\Psi_{n-}}{\pi}e^{-(\Psi^{*}_{n+}\Psi_{n+}+\Psi^{*}_{n-}\Psi_{n-})}&e^{(\Psi_{(n+1)+}^{*}\Psi_{n+}+\Psi_{n-}^{*}\Psi_{(n+1)-})}\nonumber\\& \times 
 \Big[1+\delta_t f_{\lambda}(\Psi^{*}_{(n+1)+},\Psi_{n+},\Psi^{*}_{n-},\Psi_{(n+1)-})\Big],\label{rho_n1_coh}
\end{align}
where we have used the inner product property of coherent basis i.e., $\la\Psi_{(n+1)\pm}|\Psi_{n\pm}\ra=e^{\Psi_{(n+1)\pm}^{*}\Psi_{n\pm}}$. Here $f_\lambda=f_H+f_1^{\lambda}+f_N$. Next we will go to the continuum time limit by assuming $\delta_t\rightarrow 0 $ so that we can write $\big[\Psi_{(n+1)\pm}-\Psi_{n\pm}\big]\delta_t^{-1}=\partial_t\Psi_{n \pm}$. Thus the density matrix at time $t$ in the coherent basis takes the form ,
\begin{align}
    \rho_{\lambda}(t)=\int \prod_{n=1}^{\frac{t-t_0}{\delta_t}}\frac{d\Psi_{n+}^*d\Psi_{n+}}{\pi}\frac{d\Psi_{n-}^*d\Psi_{n-}}{\pi}&e^{\delta_t[-\Psi_{n+}\partial_t\Psi^{*}_{n+}+\Psi^{*}_{n-}\partial_t\Psi_{n-}+f_\lambda(\Psi^{*}_{(n+1)+},\Psi_{n+},\Psi^{*}_{n-},\Psi_{(n+1)-})]}\nonumber\\&\quad\quad\quad\quad\quad\quad\quad\quad\times |\Psi_{(t-t_0)/\delta_t+}\ra\la\Psi_{(t-t_0)/\delta_t-}| \,\la\Psi_{0+}|\rho(t_0)|\Psi_{0-}\ra \label{eq:rho_lam_t}
\end{align}
Note that in Eq.~\eqref{eq:rho_lam_t}, $f_{\lambda}(\Psi^{*}_{(n+1)+},\Psi_{n+},\Psi_{n-}^{*},\Psi_{(n+1)-})$ has a prefactor $\delta_t$, so for $\delta_t \rightarrow 0$, we can replace $\Psi_{(n+1)\pm}$ and $\Psi^{*}_{(n+1)\pm}$ in $f_\lambda$ by $\Psi_{n\pm}$ and $\Psi_{n\pm}^{*}$ respectively using the fact that $\delta_t\Psi_{(n+1)\pm}=\delta_t\Psi_{n\pm}+O(\delta_t^2)$.
Now the MGF of the integrated current $\mathcal{N}_L$  which is defined as,
$\mathcal{Z}(\lambda)={\rm Tr} \big[\rho_{\lambda}(t_f)\big]$ (recall that $t_0$ and $t_f$ are the initial and the final time respectively)
can be written in the form of the path integral as the following,
\begin{align}
    \mathcal{Z}(\lambda)
    &=\int \prod_{n=1}^{\frac{t_f-t_0}{\delta_t}}\frac{d\Psi^{*}_{n+}d\Psi_{n+}}{\pi}\frac{d\Psi^{*}_{n-}d\Psi_{n-}}{\pi}e^{i\delta_t[\Psi_{n+}i\partial_t\Psi^{*}_{n+}-\Psi^{*}_{n-}i\partial_t\Psi_{n-}-if(\Psi^{*}_{n+},\Psi_{n+},\Psi^{*}_{n-},\Psi_{n-})]}\la\Psi_{0+}|\rho(t_0)|\Psi_{0-}\ra
    \nonumber\\&=\int D[\Psi^{*}_+,\Psi_{+},\Psi^{*}_{-},\Psi_{-}]e^{iS({\lambda})}\la\Psi_{+}(t_0)|\rho(t_0)|\Psi_{-}(t_0)\ra.\label{final_action_supp}
\end{align}
For the purpose of our analysis, as we will always deal with steady-state where the initial state is irrelevant, we ignore the term $\la\Psi_{+}(t_0)|\rho(t_0)|\Psi_{-}(t_0)\ra$ in Eq.~\eqref{final_action_supp}.
Here we identify the counting field dependent Keldysh action $S_{\lambda}$ as,
\begin{align}
    S(\lambda) = \int_{t_0}^{t_f} dt  \,\big[\Psi_{+}i\partial_t \Psi_{+}^{*}-\Psi^{*}_{-}i\partial_t \Psi_{-}]-if_{\lambda}(\Psi^{*}_{+},\Psi_{+},\Psi^{*}_{-},\Psi_{-})\big].\label{action1_supp}
\end{align}
Using integration by parts for the Grassmann variables, we finally write,
\begin{align}
    S(\lambda) = \int_{t_0}^{t_f} dt  \,\big[\Psi^{*}_{+} \big(i\partial_t \big) \Psi_{+}-\Psi^{*}_{-} \big(i\partial_t\big) \Psi_{-}]-if_{\lambda}(\Psi^{*}_{+},\Psi_{+},\Psi^{*}_{-},\Psi_{-})\big].\label{action_supp}
\end{align}
Here $f_{\lambda}(\Psi^{*}_{+},\Psi_{+},\Psi^{*}_{-},\Psi_{-})$ is given as the following,
\begin{align}
    f_{\lambda}(\Psi^{*}_{+},\Psi_{+},\Psi^{*}_{-},\Psi_{-})=&-i\sum_{i,j=1}^{N}h_{ij} [\psi_{i+}^*\psi_{j+}-\psi^*_{i-}\psi_{j-}]+\alpha_{1}\big(2e^{i\lambda}\psi^{*}_{1,+}\psi_{1,-}+\psi^{*}_{1,+}\psi_{1,+}+\psi^{*}_{1,-}\psi_{1,-}\big)\nonumber\\&+\beta_{1}\big(2e^{-i\lambda}\psi_{1,+}\psi^{*}_{1,-}-\psi^{*}_{1,+}\psi_{1,+}-\psi^{*}_{1,-}\psi_{1,-}\big)+\alpha_{N}\big(2\psi^{*}_{N,+}\psi_{N,-}+\psi^{*}_{N,+}\psi_{N,+}+\psi^{*}_{N,-}\psi_{N,-}\big)\nonumber\\&+\beta_{N}\big(2\psi_{N,+}\psi^{*}_{N,-}-\psi^{*}_{N,+}\psi_{N,+}-\psi^{*}_{N,-}\psi_{N,-}\big).
    \label{supp-f-lambda}
\end{align}
Eq.~\eqref{action_supp} along with Eq.~\eqref{supp-f-lambda} is the appropriate action, as given in the main text in Eq.~\eqref{action1}. 
\section{Derivation of the Keldysh action in Eq.~\eqref{eq:action_gainloss} for case-2: boundary drives + gain and loss channels in the bulk}
\label{sec:app2}
In this appendix, we derive the Keldysh action in the presence of gain and loss channels in the bulk along with boundary drives. In this case, the dressed Liouvillian $\mathcal{L}_\lambda\bullet$ in Eq.~\eqref{liou_boundry} has an extra dissipative contribution due to the gain-loss channels which is given by,
\begin{align}
    \mathcal{L}_\lambda\,\bullet=-iH\bullet + \bullet iH+\mathcal{D}^{\lambda}_1[\bullet]+\mathcal{D}_N[\bullet]+\mathcal{D}_{\rm bulk}[\bullet], \label{liou_gain_loss_supp}
\end{align}
where recall that
\begin{align}\mathcal{D}^{\lambda}_1[\bullet]=\alpha_{1}\big[2e^{i\lambda} c^{\dagger}_{1}\bullet c_{1}-\{c_{1}c^{\dagger}_{1},\bullet\} \big]+\beta_{1}\big[2e^{-i\lambda}c_{1}\bullet c^{\dagger}_{1}-\{c^{\dagger}_{1}c_{1},\bullet\} \big],
\end{align}
\begin{align}
\mathcal{D}_{N}[\bullet]=\alpha_{N}\big[2c^{\dagger}_{N}\bullet c_{N}-\{c_{N}c^{\dagger}_{N},\bullet\} \big]+\beta_{N}\big[2 c_{N}\bullet c^{\dagger}_{N}-\{c^{\dagger}_{N}c_{N},\bullet\} \big],
\end{align}
\begin{align}
\mathcal{D}_{\rm bulk}[\bullet]=&\sum_{i=1}^{N-1}\Big(
\kappa_1 [2c_{i}\bullet c_{i}^{\dagger}\!-\!\{c^{\dagger}_{i}c_i,\bullet\}]
+\kappa_2[2c_{i+1}\bullet c_{i+1}^{\dagger}\!-\!\{c^{\dagger}_{i+1}c_{i+1},\bullet\}]
\nonumber\\
&\quad
+e^{-i\phi}\sqrt{\kappa_1\kappa_2}\,[2c_{i}\bullet c_{i+1}^{\dagger}\!-\!\{c^{\dagger}_{i+1}c_{i},\bullet\}]
+e^{i\phi}\sqrt{\kappa_1\kappa_2}\,[2c_{i+1}\bullet c_{i}^{\dagger}\!-\!\{c^{\dagger}_{i}c_{i+1},\bullet\}]
\Big)
\nonumber\\
&+\sum_{i=1}^{N-1}\Big(
\Gamma_1 [2c^{\dagger}_{i}\bullet c_{i}\!-\!\{c_{i}c^{\dagger}_i,\bullet\}]
+\Gamma_2[2c^{\dagger}_{i+1}\bullet c_{i+1}\!-\!\{c_{i+1}c^{\dagger}_{i+1},\bullet\}]
\nonumber\\
&\quad
+e^{-i\theta}\sqrt{\Gamma_1\Gamma_2}\,[2c^{\dagger}_{i}\bullet c_{i+1}\!-\!\{c_{i+1}c^{\dagger}_{i},\bullet\}]
+e^{i\theta}\sqrt{\Gamma_1\Gamma_2}\,[2c^{\dagger}_{i+1}\bullet c_{i}\!-\!\{c_{i}c^{\dagger}_{i+1},\bullet\}]
\Big)
\end{align}
To obtain the coherent basis representation of $\rho^{\lambda}_{n+1}$ for this case, we follow the same steps from Eq.~\eqref{rho_n_supp}-\eqref{r_n1_supp} used in the previous appendix. 
Recall that, using $\rho^{\lambda}_{n+1}=[\mathbb{I}+\delta_t\mathcal{L}_\lambda]\rho^{\lambda}_n$ and insert the resolution of identity in terms of coherent states in the left and right side of $\rho^{\lambda}_n$, we can write,
\begin{align}
\rho_{n+1}^{\lambda}=\int& \frac{d\Psi^{*}_{n+}d\Psi_{n+}}{\pi}\frac{d\Psi^{*}_{n-}d\Psi_{n-}}{\pi}e^{-(\Psi^{*}_{n+}\Psi_{n+}+\Psi^{*}_{n-}\Psi_{n-})}|\Psi_{n+}\rangle\la\Psi_{n-}|\la\Psi_{n+}|\rho_n^{\lambda}|\Psi_{n-}\ra\nonumber\\&+\delta_t\int \frac{d\Psi^{*}_{n+}d\Psi_{n+}}{\pi}\frac{d\Psi^{*}_{n-}d\Psi_{n-}}{\pi}e^{-(\Psi^{*}_{n+}\Psi_{n+}+\Psi^{*}_{n-}\Psi_{n-})}\mathcal{L}_{\lambda}\Big[|\Psi_{n+}\rangle\la\Psi_{n-}|\Big]\la\Psi_{n+}|\rho_n^{\lambda}|\Psi_{n-}\ra. 
\end{align}
Next task is to obtain $\la\Psi_{(n+1)+}|\mathcal{L}_\lambda[|\Psi_{n+}\ra\la\Psi_{n-}|]|\Psi_{(n+1)-}\ra$ for the Liouvillian given in Eq.~\eqref{liou_gain_loss_supp}. Similar to the Eq.~\eqref{liouvel_coherent_supp}, here we obtain,
\begin{align}           \la\Psi_{(n+1)+}&|\mathcal{L}_\lambda[|\Psi_{n+}\ra\la\Psi_{n-}|]|\Psi_{(n+1)-}\ra=\Big[f_H(\Psi^{*}_{(n+1)+},\Psi_{n+},\Psi^{*}_{n-},\Psi_{(n+1)-})+f_1(\Psi^{*}_{(n+1)+},\Psi_{n+},\Psi^{*}_{n-},\Psi_{(n+1)-})\nonumber\\&+f_N(\Psi^{*}_{(n+1)+},\Psi_{n+},\Psi^{*}_{n-},\Psi_{(n+1)-})+f_{\rm bulk}(\Psi^{*}_{(n+1)+},\Psi_{n+},\Psi^{*}_{n-},\Psi_{(n+1)-})\Big]\la\Psi_{(n+1)+}|\Psi_{n+}\ra\la\Psi_{n-}|\Psi_{(n+1)-}\ra,\label{liouvel_coherent_gainloss_supp}
\end{align}
where the Hamiltonian contribution, $f_H(\Psi^{*}_{(n+1)+},\Psi_{n+},\Psi^{*}_{n-},\Psi_{(n+1)-})$, the contribution of the boundary dissipators, $f_1^{\lambda}(\Psi^{*}_{(n+1)+},\Psi_{n+},\Psi^{*}_{n-},\Psi_{(n+1)-})$, and $f_N(\Psi^{*}_{(n+1)+},\Psi_{n+},\Psi^{*}_{n-},\Psi_{(n+1)-})$ are same as Eq.~\eqref{fH_supp_boundary}, \eqref{f1_supp} and \eqref{fN_supp} respectively, which is obtained in  the previous Appendix~\ref{app:action}.
The contribution of the bulk dissipators, $f_{\rm bulk}(\Psi^{*}_{(n+1)+},\Psi_{n+},\Psi^{*}_{n-},\Psi_{(n+1)-})$ is obtained as,
\begin{align}
    &f_{\rm bulk}=\sum_{i=1}^{N-1}\Bigg\{\Gamma_1\big[2\psi^{i}_{(n+1)+}\psi^{i*}_{(n+1)-}\!+\!\psi_{(n+1)+}^{i*}\psi_{n+}^{i}\!+\!\psi_{n-}^{i*}\psi_{(n+1)-}^{i}\!\big]\!+\!\Gamma_2\big[2\psi^{i+1}_{(n+1)+}\psi^{i+1*}_{(n+1)-}\!+\!\psi_{(n+1)+}^{i+1*}\psi_{n+}^{i+1}\!+\!\psi_{n-}^{i+1*}\psi_{(n+1)-}^{i+1}\!\big]\nonumber\\&+e^{-i\theta}\sqrt{\Gamma_1\Gamma_2}\big[2\psi^{i}_{(n+1)+}\psi^{i+1*}_{(n+1)-}+\psi_{(n+1)+}^{i+1*}\psi_{n+}^{i}+\psi_{n-}^{i+1*}\psi_{(n+1)-}^{i}\big]+e^{i\theta}\sqrt{\Gamma_1\Gamma_2}\big[2\psi^{i+1}_{(n+1)+}\psi^{i*}_{(n+1)-}+\psi_{(n+1)+}^{i*}\psi_{n+}^{i+1}+\psi_{n-}^{i*}\psi_{(n+1)-}^{i+1}\big]\nonumber\\&+\Big(\kappa_1\big[2\psi^{i*}_{n+}\psi^{i}_{n-}-\psi_{(n+1)+}^{i*}\psi_{n+}^{i}-\psi_{n-}^{i*}\psi_{(n+1)-}^{i}\big]+\kappa_2\big[2\psi^{i+1*}_{n+}\psi^{i+1}_{n-}-\psi_{(n+1)+}^{i+1*}\psi_{n+}^{i+1}-\psi_{n-}^{i+1*}\psi_{(n+1)-}^{i+1}\big]\nonumber\\&+e^{-i\phi}\sqrt{\kappa_1\kappa_2}\big[2\psi^{i*}_{n+}\psi^{i+1}_{n-}-\psi_{(n+1)+}^{i+1*}\psi_{n+}^{i}-\psi_{n-}^{i+1*}\psi_{(n+1)-}^{i}\big]+e^{i\phi}\sqrt{\kappa_1\kappa_2}\big[2\psi^{i+1*}_{n+}\psi^{i}_{n-}-\psi_{(n+1)+}^{i*}\psi_{n+}^{i+1}-\psi_{n-}^{i*}\psi_{(n+1)-}^{i+1}\big]\Big)\Bigg\}\label{f_bulk_supp}
\end{align}
Thus $f_\lambda(\Psi^{*}_{n+},\Psi_{n+},\Psi^{*}_{n-},\Psi_{n-})=f_H+f_1^\lambda+f_N+f_{\rm bulk}$ and then following similar steps as done in Eq.~\eqref{rho_n1_coh}-\eqref{action_supp}, we obtain the action,
\begin{align}
S(\lambda)=\int_{t_0}^{t_f}dt[\Psi^{*}_{+} \big(i\partial_t \big) \Psi_{+}-\Psi^{*}_{-} \big(i\partial_t\big)\Psi_{-}-if_\lambda(\Psi^{*}_{+},\Psi_{+},\Psi^{*}_{-},\Psi_{
    -})],
\end{align}
\twocolumngrid
\bibliography{references.bib}
    
\end{document}